\documentclass{emulateapj}
\usepackage{natbib,apjfonts}
\newcommand{\HI}{\ion{H}{1}}

\newcommand{\Msol}{$M_\odot$}
\newcommand{\kms}{\mbox{km~s$^{-1}$}}

\shorttitle{Molecular and Atomic Gas in the LMC. I.}
\submitted{Received 17 Nov 2008; Accepted 9 Feb 2009}
\journalinfo{To Appear in the Astrophysical Journal}

\begin{document}

\title{Molecular and Atomic Gas in the Large Magellanic Cloud - I. Conditions for CO Detection}

\author{T.~Wong\altaffilmark{1,2}, A.~Hughes\altaffilmark{3,2}, Y.~Fukui\altaffilmark{4}, A.~Kawamura\altaffilmark{4}, N.~Mizuno\altaffilmark{4,5}, J.~Ott\altaffilmark{6,7,8}, E.~Muller\altaffilmark{4,2}, J.~L.~Pineda\altaffilmark{9,10}, D.~E.~Welty\altaffilmark{1},
S.~Kim\altaffilmark{11}, Y.~Mizuno\altaffilmark{4}, M.~Murai\altaffilmark{4}, \& T. Onishi\altaffilmark{4}}

\altaffiltext{1}{Astronomy Department, University of Illinois, 1002 W. Green St, Urbana, IL 61801; wongt@astro.uiuc.edu.}
\altaffiltext{2}{CSIRO Australia Telescope National Facility, PO Box 76, Epping, NSW 1710, Australia.}
\altaffiltext{3}{Centre for Astrophysics and Supercomputing, Swinburne University of Technology, PO Box 218, Hawthorn, VIC 3122, Australia.}
\altaffiltext{4}{Department of Physics, Nagoya University, Chikusa-ku, Nagoya 464-8602, Japan.}
\altaffiltext{5}{ALMA-J Project Office, National Astronomical Observatory of Japan, 2-21-1 Osawa, Mitaka, 181-8588 Tokyo, Japan.}
\altaffiltext{6}{National Radio Astronomy Observatory, 520 Edgemont Rd, Charlottesville, VA 22903.}
\altaffiltext{7}{California Institute of Technology, 1200 E. California Blvd, Caltech Astronomy, 105-24, Pasadena, CA 91125-2400.}
\altaffiltext{8}{Jansky Fellow of the National Radio Astronomy Observatory.}
\altaffiltext{9}{Jet Propulsion Laboratory, California Institute of Technology, 4800 Oak Grove Drive, Pasadena, CA 91109.}
\altaffiltext{10}{NASA Postdoctoral Program Fellow.}
\altaffiltext{11}{Department of Astronomy and Space Science, Sejong University, KwangJin-gu, KunJa-dong 98, Seoul 143-747, Korea.}

\begin{abstract}

We analyze the conditions for detection of CO(1-0) emission in the Large Magellanic Cloud (LMC), using the recently completed second NANTEN CO survey.  In particular, we investigate correlations between CO integrated intensity and \HI\ integrated intensity, peak brightness temperature, and line width at a resolution of 2\farcm6 ($\sim$40 pc).  We find that significant \HI\ column density (exceeding $\sim$$10^{21}$ cm$^{-2}$) and peak brightness temperature (exceeding $\sim$20 K) are necessary but not sufficient conditions for CO detection, with many regions of strong \HI\ emission not associated with molecular clouds.  The large scatter in CO intensities for a given \HI\ intensity persists even when averaging on scales of $>$200 pc, indicating that the scatter is not solely due to local conversion of \HI\ into H$_2$ near GMCs.  We focus on two possibilities to account for this scatter: either there exist spatial variations in the $I$(CO) to $N$(H$_2$) conversion factor, or a significant fraction of the atomic gas is not involved in molecular cloud formation.  A weak tendency for CO emission to be suppressed for large \HI\ linewidths supports the second hypothesis, insofar as large linewidths may be indicative of warm \HI, and calls into question the likelihood of forming molecular clouds from colliding \HI\ flows.  We also find that the ratio of molecular to atomic gas shows no significant correlation (or anti-correlation) with the stellar surface density, though a correlation with midplane hydrostatic pressure $P_h$ is found when the data are binned in $P_h$.  The latter correlation largely reflects the increasing likelihood of CO detection at high \HI\ column density.

\end{abstract}

\keywords{Magellanic Clouds---ISM: molecules---galaxies: ISM}

\section{Introduction}\label{sec:intro}

A key step in the cycling of gas between stars and the interstellar medium is the formation of molecular clouds, which in turn serve as the sites of further star formation.  Understanding this process requires observations of both the atomic gas component, observed most readily using the 21-cm line of \HI, and the molecular gas component, observed most readily using the 2.6-mm $J$=1$\rightarrow$0 line of CO.  While the information provided by these two spectral lines is not unambiguous---both, especially CO, can be affected by optical depth and excitation effects---they provide rough indications of gas column densities, mean velocities, and velocity dispersions which can be used to constrain cloud formation models.

Balancing H$_2$ formation on grains and photodissociation by far-ultraviolet (FUV) radiation leads one to expect a core-envelope structure for molecular clouds, in which dense CO-emitting cores are shielded by \HI\ envelopes.  The tendency for \HI\ mass surface densities to saturate at $\Sigma \sim 10$ \Msol\ pc$^{-2}$ or $N_{\rm H} \sim 10^{21}$ cm$^{-2}$ \citep[e.g.,][]{Wong:02} suggests that much of the \HI\ gas in galaxies may consist of such envelopes, with any additional gas being transformed into H$_2$ \citep{Shaya:87}.  Direct evidence for \HI\ envelopes around CO clouds has been presented by \citet{Wannier:83} and \citet{Andersson:91}, who found \HI\ emission maxima near the edges of CO clouds.  More recently, \citet{Goldsmith:05} and \citet{Goldsmith:07} have used time-dependent models of H$_2$ formation, in conjunction with observations of \HI\ self-absorption in molecular clouds, to show that compressed \HI\ clouds become molecular from the inside out; the clouds they study have inferred minimum ages of 3--10 Myr.  Since the interstellar medium (ISM) is thermally bistable \citep{Field:69}, the raw material for molecular cloud formation is expected to be the cold neutral medium (CNM) traced by 21-cm absorption studies, which occupies a much smaller volume than the low-density warm neutral medium (WNM).

While molecular clouds are generally assumed to be quasi-equilibrium structures, a more dynamic view of the interstellar medium has recently been gaining prominence.  In this picture, shocks driven by large-scale motions are the dominant process for producing density enhancements, and molecular clouds can form rapidly downstream of shock fronts \citep[e.g.,][]{Bergin:04,Glover:07b}, a scenario consistent with the relatively short inferred lifetimes for molecular clouds in the solar neighborhood \citep{Hartmann:01}.  If the timescale on which gas experiences shocks is shorter than the timescale for thermal instability to operate, then the CNM may not be a necessary precursor to molecular clouds, and one can even envision molecular clouds forming directly from shocked WNM gas \citep[e.g.,][]{Bergin:04}.  We note, however, that the simulations of \citet{Glover:07b}, which form H$_2$ on timescales of only $\sim$1 Myr, begin with gas already at densities typical of the CNM ($\sim$100 cm$^{-3}$).

It is unclear how rapid molecular cloud formation would affect the observed relationship between \HI\ and CO.\@  \HI\ would still be a necessary ingredient for cloud formation, and indeed some turbulent cloud formation models predict a long ``latency'' period of \HI\ accumulation prior to a molecular cloud becoming observable in CO \citep{Vazquez:07}.  Molecular clouds formed recently from colliding gas flows should be associated with larger \HI\ line widths, but even this observational signature may be obscured if gravitational collapse is delayed while \HI\ flows are active \citep{Vazquez:07}, or if the \HI\ line width is dominated by the WNM, whose thermal velocity dispersion exceeds the turbulent velocity dispersion of the newly-formed CNM \citep{Koyama:02}.  Another possible signature of turbulence may be the existence of a porous density structure with significant mixing of atomic and molecular material \citep[e.g.,][]{Hennebelle:08}, thus tending to erase any sharp column density threshold for H$_2$ formation associated with self-shielding.

The Large Magellanic Cloud (LMC) is the nearest star-forming galaxy to our own, and its nearly face-on aspect makes it an excellent target for studying the relationships between atomic and molecular clouds and star formation.  In this study we compare the global CO and \HI\ distributions at 2\farcm6 resolution, corresponding to a spatial scale of $\sim$40 pc (assuming a distance of 50 kpc to the LMC).  While this scale resolves only the largest molecular clouds, it is comparable to the linear dimension of the area from which atomic gas must be gathered in order to form a giant molecular cloud of mass $10^4$--$10^5$ \Msol.  We consider the detectability of CO emission and its dependence on \HI\ column density, peak brightness, and velocity dispersion.  We also evaluate the coincidence between CO and \HI\ emission in velocity space using a Gaussian decomposition method.  Finally, we consider whether simple estimates of the hydrostatic disk pressure are able to predict the location of regions with high molecular to atomic gas ratios.

Previous comparisons of CO and \HI\ emission in the LMC have focused on specific regions or been undertaken at spatial scales much larger than 3\arcmin.  The only global comparison, aside from a brief discussion in \citet{Blitz:07}, was made by \citet{Mizuno:01}, who focused on the association of the large-scale ``L'' and ``D'' components of the galaxy seen in single-dish \HI\ maps (15\arcmin\ beam FWHM) with the CO clouds mapped in the first NANTEN survey.  In a companion paper (Fukui et al.\ 2009, hereafter Paper II), we will examine the extent of the HI emission (in position and velocity space) around CO-detected clouds, and relate the CO and \HI\ properties of giant molecular clouds with their star formation activity.

\begin{figure*}
\begin{center}
\includegraphics[height=\textwidth,angle=-90]{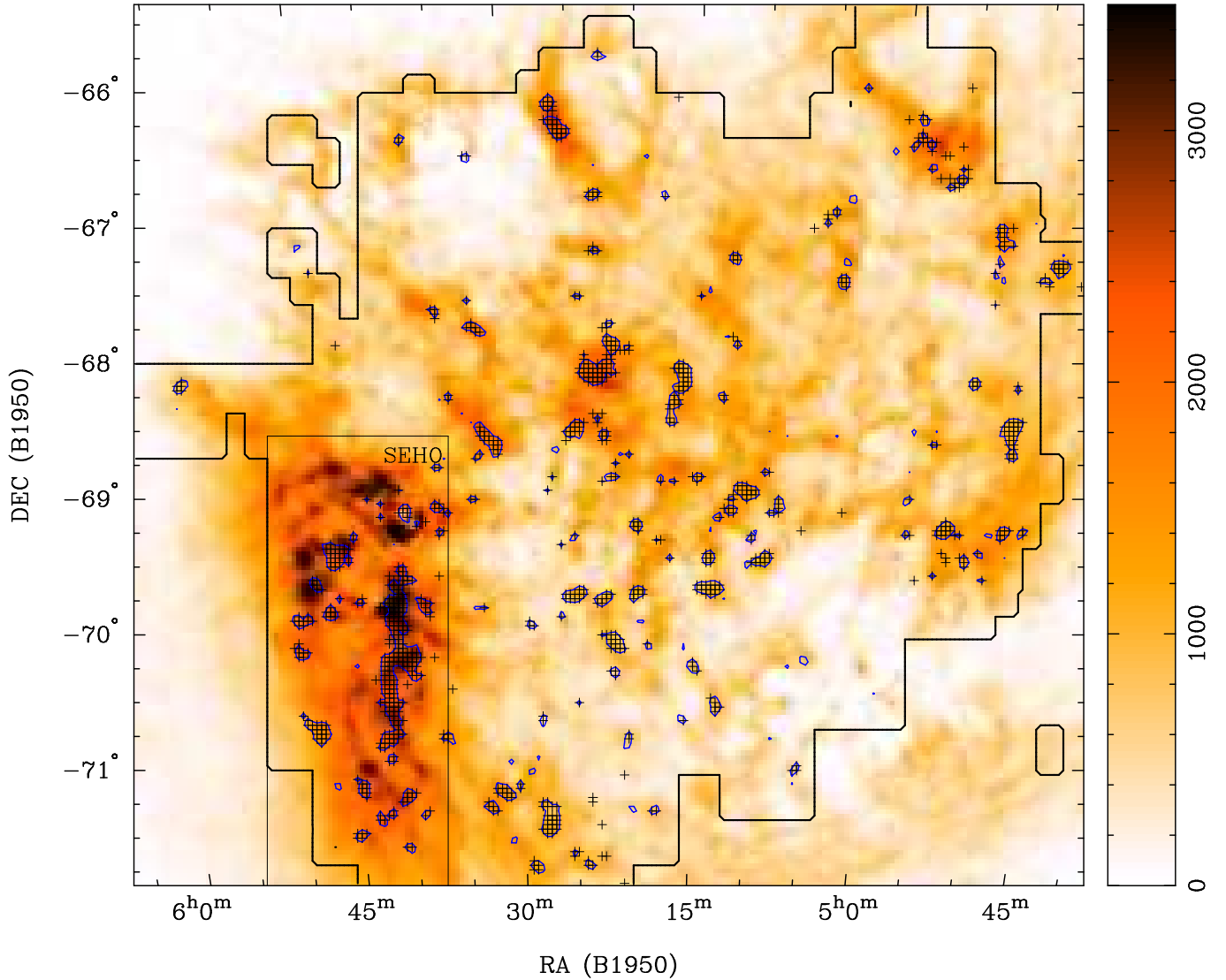}
\end{center}
\caption{
CO contours from NANTEN (at the 1 K \kms\ level) overlaid on integrated \HI\ emission smoothed to the resolution of the CO data.  Units of the \HI\ intensity are also K \kms.  To reject noise in the contour map, a blanking mask constructed with the CPROPS signal detection algorithm described in \S\ref{sec:gaufit} has been applied.  The heavy solid contour represents the region observed with NANTEN.  Small crosses indicate the pixels which are considered CO detections without the use of a blanking mask.  The southeastern \HI\ overdensity (SEHO) is identified at the lower left of the map.
\label{fig:maps}}
\end{figure*}

\section{Observational Data}\label{sec:obs}

The \HI\ data for the LMC were taken with the Australia Telescope Compact Array (ATCA) and the Parkes telescopes, both part of the Australia Telescope National Facility.\footnote{The Australia Telescope is funded by the Commonwealth of Australia for operation as a National Facility managed by CSIRO.}  Details of the observations are given by \citet{Kim:98} and \citet{Staveley-Smith:03}, and the data combination is described by \citet{Kim:03}.  The resulting data cube covers a region of 7\fdg5 $\times$ 7\fdg5 at a spatial resolution of 1\arcmin\ and with a 1$\sigma$ noise level of 2.4~K in a 1.65 \kms\ channel.  To facilitate comparison with the CO data, we regridded the \HI\ spectra to the LSR velocity frame and smoothed the maps to a resolution of 2\farcm6; the resulting RMS noise in the channel maps was 0.6~K per channel.  At this resolution the maximum observed \HI\ brightness temperature is $\sim$110 K, so the peak signal-to-noise ratio is $\sim$190.

The CO data were taken with NANTEN, a 4-m radio telescope operated by Nagoya University at Las Campanas Observatory in Chile.  A region of $\sim$30 square degrees was observed in 26,877 positions with a 2\arcmin\ grid spacing, slightly undersampling the telescope primary beam (HPBW=2\farcm6).  Each position was observed for about 3 minutes on-source, yielding a typical 1$\sigma$ RMS noise of 0.07 K in a 0.65 \kms\ channel.  The maximum observed CO brightness was 2.1 K, corresponding to a peak signal-to-noise ratio of $\sim$30.  A cube was formed from the spectra using the GLS projection in 1950.0 coordinates from 150 to 350 \kms\ in the $v_{\rm LSR}$ frame.  However, spectra in some regions of the galaxy only cover 100 \kms\ in velocity, generally centered near the peak \HI\ velocity.  Further details about the observations and data processing can be found in \citet{Fukui:08}.

Since different parts of the LMC were observed to different sensitivity limits in CO, we constructed a noise image using the median absolute deviation (MAD), considering only channels in each spectrum with $T_b < 0.5$~K.  The MAD, which is more resistant to outliers than the standard deviation, is obtained by calculating the absolute values of the residuals (deviations) from the median of the spectrum, then taking the median.  The 1$\sigma$ noise estimate is then 1.48 $\times$ MAD.  Obvious rectangular patches in the noise image indicate that large regions of the galaxy were observed to comparable sensitivity, presumably during continuous observing sessions.  To emphasize these large scale noise variations (which span more than a factor of 2, from 0.045 to 0.1 K) and reduce the intrinsic scatter when deriving noise estimates from individual spectra, we applied a moving 3 $\times$ 3 pixel median filter to the noise image using the MFILTER task in GIPSY.

The integrated \HI\ and CO intensities were derived straightforwardly by summing the intensity for all spectral channels at a given position.  \citet{Fukui:08} found that all of the detected CO emission was restricted to a $v_{\rm LSR}$ range of 210--310 \kms.  To reduce the noise contribution to the integrated CO map, we only summed the CO emission from 200--325 \kms, although we summed the \HI\ emission from 175--350 \kms\ in order to accommodate the typically broader \HI\ linewidths ($\sigma_v \approx 12$ \kms) compared to CO ($\sigma_v \approx 3$ \kms).  Figure~\ref{fig:maps} shows the resulting integrated \HI\ image.  The thick solid contour represents the region over which CO emission has been observed.  For most of our analysis we have avoided identifying and blanking emission-free regions, so as to provide a simple estimate of the sensitivity at each location.  However, for purposes of overlaying a fiducial 1 K \kms\ CO contour in Figure~\ref{fig:maps} we have used a signal detection algorithm (discussed in \S\ref{sec:gaufit} below) to only sum channels which are adjacent to, or coincide with, 3$\sigma$ detections.  Visual comparison between the resulting CO contour and the distribution of pixels detected in integrated intensity without blanking (shown as crosses) indicates that they are largely consistent.

\begin{figure*}
\centerline{\includegraphics[width=0.8\textwidth,bb=32 170 575 695]{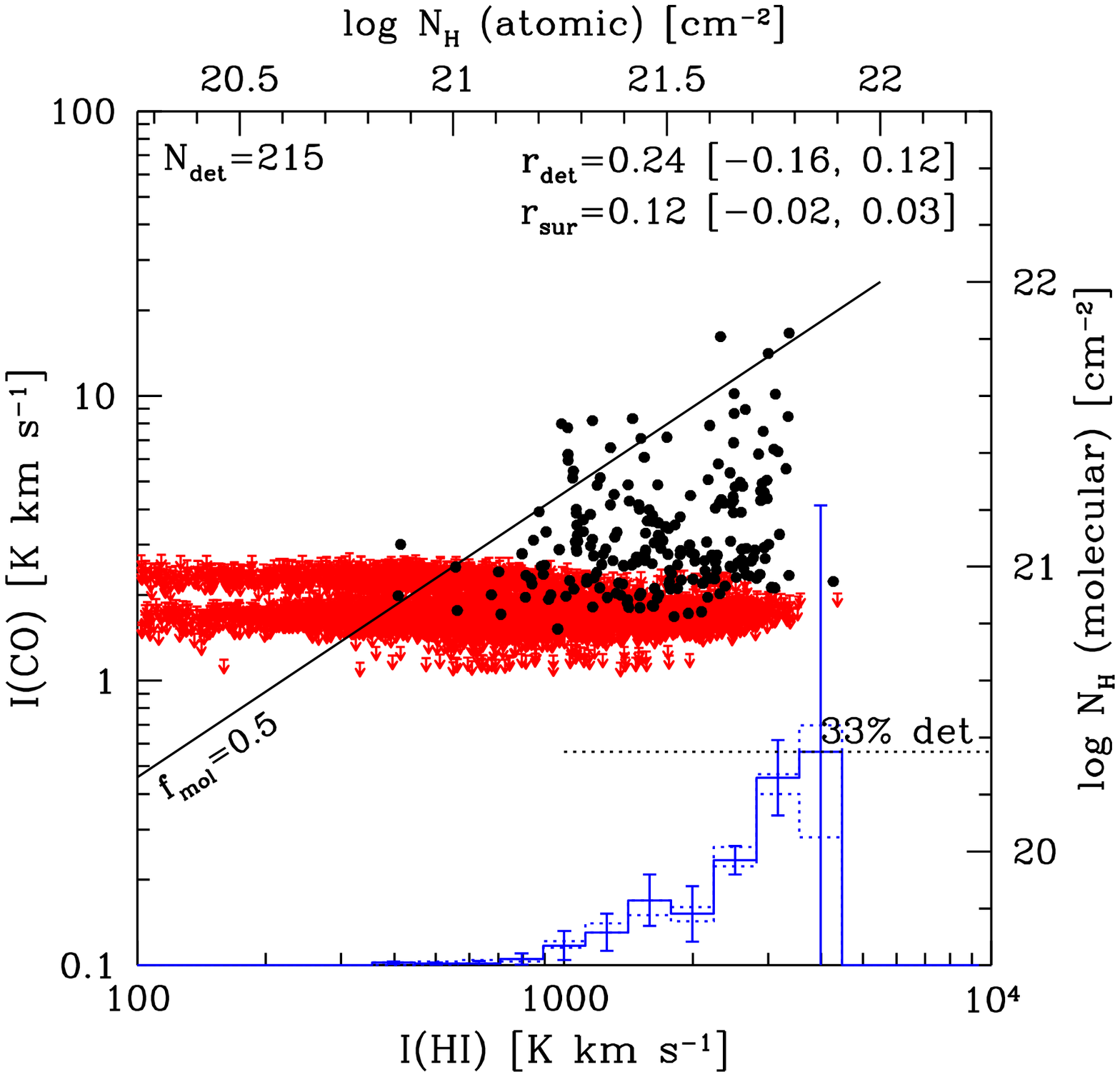}}
\caption{
Integrated CO intensity vs.\ \HI\ intensity for 2\arcmin\ $\times$ 2\arcmin\ pixels separated by 4\arcmin.  Black solid circles are CO detections, red arrows represent 3$\sigma$ upper limits.  At the bottom is a histogram of the CO detection fraction in equally spaced logarithmic bins of $I_{\rm HI}$, with a linear ordinate peaking at the value indicated.  Since only about a quarter of the pixels are shown, dashed histograms indicate the range of values obtained from the other three sets of pixels.  The diagonal solid line indicates a molecular mass fraction of 0.5, assuming optically thin \HI\ and a Galactic value of the CO-to-H$_2$ conversion factor.  Correlation coefficients are explained in the text.
\label{fig:corrmom}}
\end{figure*}

\section{Results}

\subsection{CO Intensity vs.\ \HI\ Properties}\label{results:props}

In this section we consider the detectability of CO emission as a function of \HI\ peak brightness temperature (hereafter $T_{\rm max}$), linewidth, and integrated intensity.  We phrase the comparison in this way because the limiting factor in comparing the two species is the ability to detect CO: not only is the peak signal-to-noise ratio a factor of $\sim$6 higher for the \HI\ data, but the fraction of spectra with $T_{\rm max}$ exceeding 50\% of the global $T_{\rm max}$ is much higher for \HI\ (8.8\%) than for CO (0.15\%), implying that extremely few regions in the galaxy are bright in CO.

The comparison between CO and \HI\ can be done in 2-D (using e.g., integrated intensity maps) or in 3-D (as a function of radial velocity as well as position).  Each has advantages and disadvantages: a 2-D analysis can lead one to compare unrelated material as a result of line-of-sight superposition, whereas a 3-D analysis can weaken an underlying correlation if there are differences in intrinsic linewidths resulting from differences in temperature or optical depth.  In this paper our main approach is a 2-D one, but in \S\ref{sec:gaufit} we employ multiple Gaussian fits to better isolate the relevant \HI\ component.  We defer a full 3-D analysis of the data cubes to Paper II.

For the present analysis, we select every second pixel along each of the spatial axes, to ensure each pixel is statistically independent (pixel separation 4\arcmin).  Figure~\ref{fig:corrmom} compares the integrated \HI\ and CO intensities, with pixels detected in both \HI\ and CO shown as filled circles, and pixels detected only in \HI\ shown as upper limit symbols (the 3$\sigma$ sensitivity limit for detecting \HI, $I_{\rm HI}$$\approx$33 K~\kms, lies beyond the left edge of the panel).  On the top and right plot axes we have converted the integrated intensities into equivalent hydrogen nucleon column densities, assuming optically thin \HI\ and a CO-to-H$_2$ conversion factor appropriate to the Galaxy, $N({\rm H_2})/I_{\rm CO}$=$2 \times 10^{20}$ cm$^{-2}$ (K \kms)$^{-1}$ \citep{Strong:96}.  Upper limit symbols correspond to 3$\sigma$ upper limits for the integrated CO intensity.  Of approximately 6700 pixels in the region observed by NANTEN, only 215 (3.2\%) are detected in both tracers, while only 22 (0.3\%) are not detected in either tracer.  Only 1 pixel is detected in CO but not \HI, corresponding to the bright continuum source 30 Doradus where \HI\ is seen in absorption.  Thus, over most of the area surveyed by NANTEN, \HI\ is detected but not CO.  Below we discuss the detectability of CO as a function of several characteristic properties of the \HI\ spectra.

\begin{figure*}
\includegraphics[width=8cm]{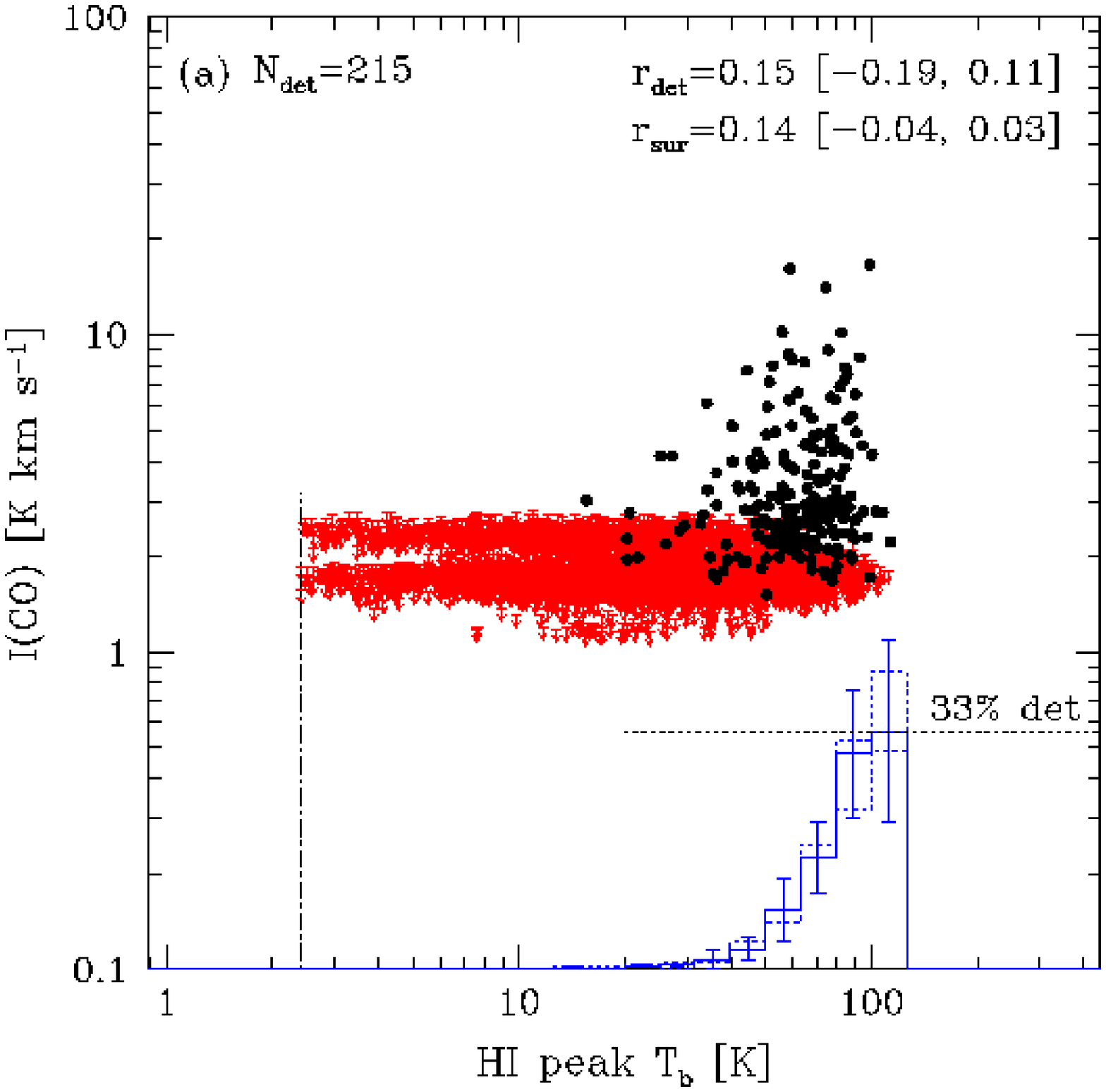}\hfill
\includegraphics[width=8cm]{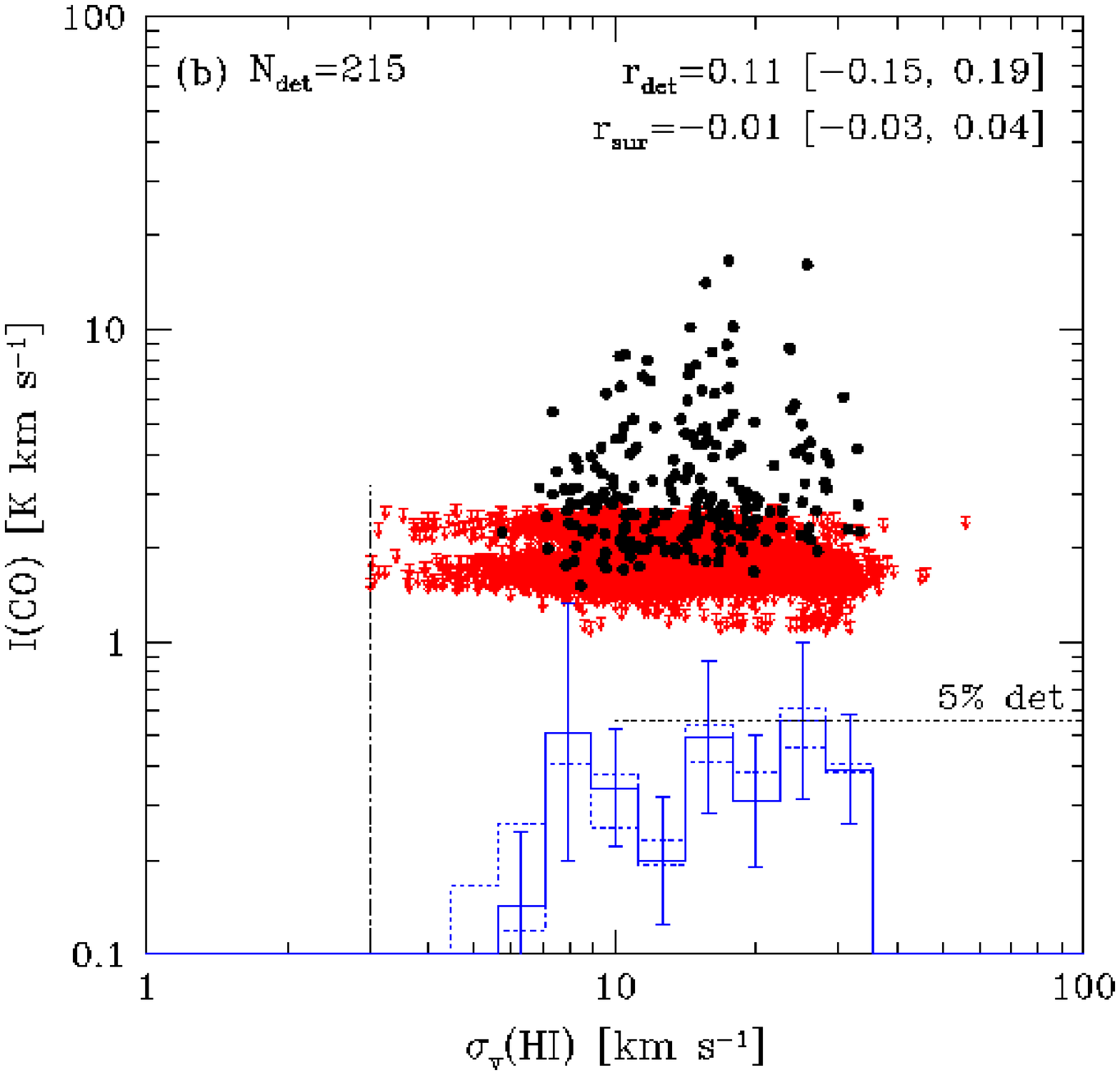}
\caption{\footnotesize
(a) Integrated CO intensity vs.\ \HI\ peak brightness temperature for 2\arcmin\ $\times$ 2\arcmin\ pixels separated by 4\arcmin.   (b) Integrated CO intensity vs.\ \HI\ velocity dispersion.  For both panels, symbols and histograms are as in Figure~\ref{fig:corrmom}.  Values to the left of the vertical dot-dashed lines are below our adopted sensitivity limits and thus excluded. 
\label{fig:corrpk}}
\end{figure*}

\begin{figure*}
\includegraphics[width=8cm,bb=32 156 575 680]{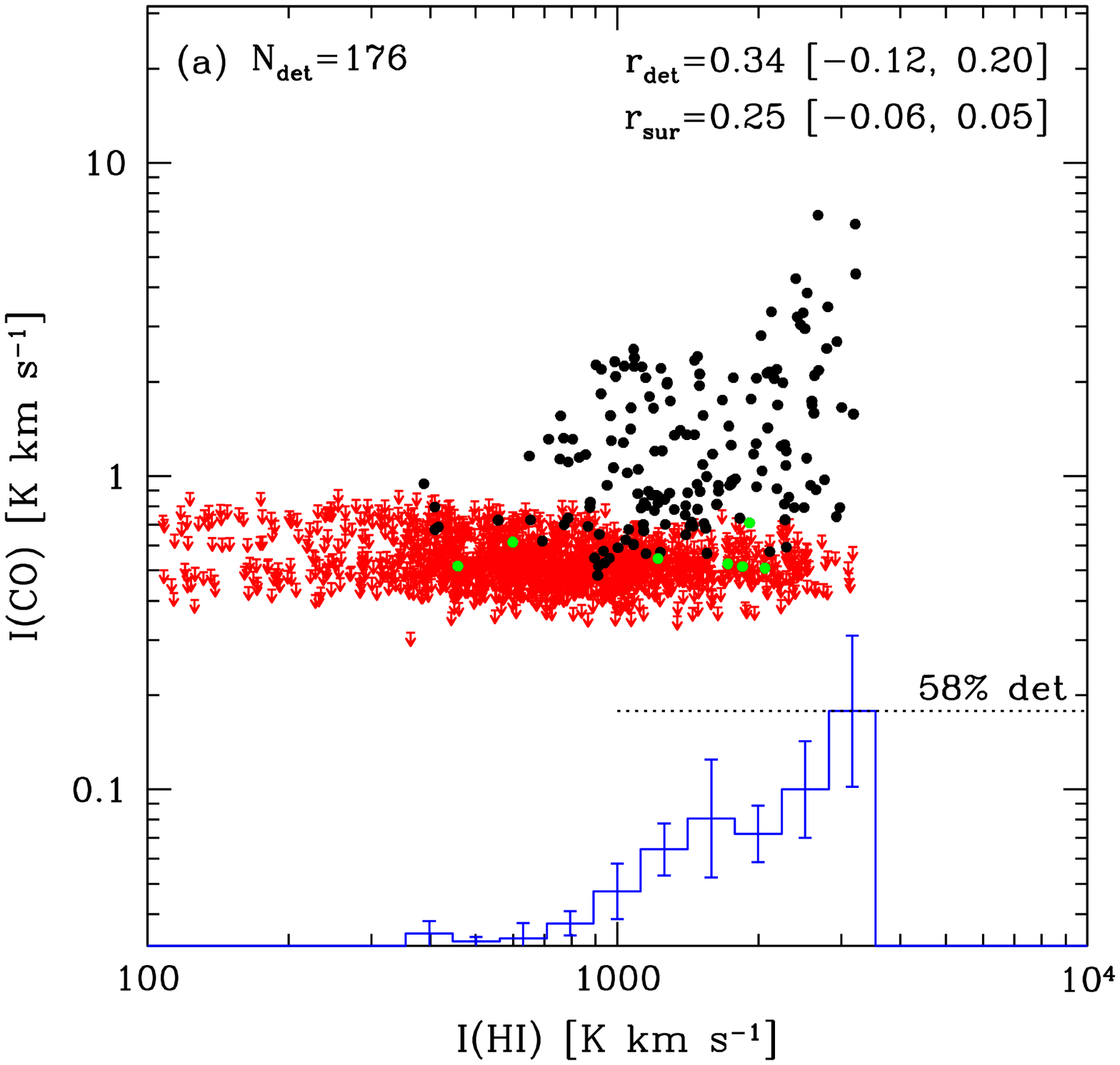}\hfill
\includegraphics[width=8cm,bb=32 156 575 680]{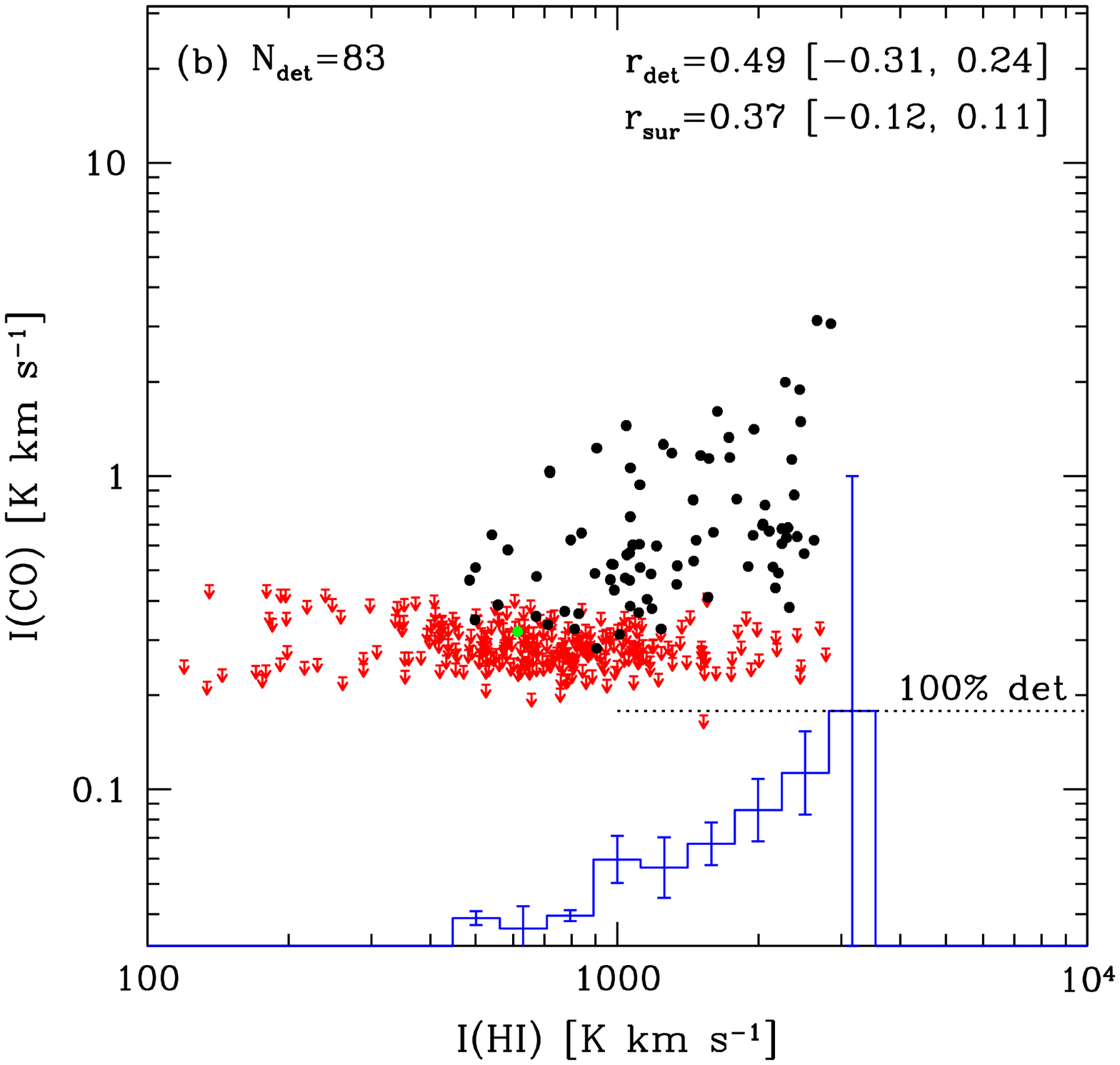}
\caption{
(a) Integrated CO intensity vs.\ \HI\ intensity for 8\arcmin\ $\times$ 8\arcmin\ pixels.
(b) Integrated CO intensity vs.\ \HI\ intensity for 16\arcmin\ $\times$ 16\arcmin\ pixels.  For both panels, symbols and histograms are as in Figure~\ref{fig:corrmom}, except that green solid circles indicate pixels which are classified as detections even though none of the constituent 2\arcmin\ pixels were considered detections.
\label{fig:bin48}}
\end{figure*}

\subsubsection{\HI\ Integrated Intensity}

Figure~\ref{fig:corrmom} illustrates the tendency for strong CO emission (as measured by the integrated line intensity) to be associated with strong \HI\ emission.  The correlation is by no means perfect: many regions with strong \HI\ emission show little or no CO emission.  The Spearman rank correlation coefficient, calculated for points detected in both tracers, is only $r_{\rm det}=0.24$, and the generalized rank correlation coefficient, taking into account CO upper limits using a survival analysis \citep{Isobe:86}, is only $r_{\rm surv}=0.12$.  Both correlations nonetheless appear to be larger than would be obtained by chance, as each lies outside the range of values given in brackets, which represents the full range of correlation coefficients obtained when randomly shuffling the abscissas of the data points 50 times.

A histogram of the CO detection fraction is shown at the bottom of the plot.  Placing the pixels into logarithmic bins of increasing \HI\ intensity, the fraction of pixels with detected CO increases monotonically, peaking around 33\% (although the last bin suffers from small number statistics).  It is noteworthy that weak \HI\ emission (below an integrated intensity $I_{\rm HI} \approx 700$ K \kms, or $N_{\rm H} \approx 10^{21}$ cm$^{-2}$) is never associated with strong CO emission (above an integrated intensity $I_{\rm CO} \approx 3$ K \kms).  Thus, high \HI\ intensities are {\it necessary but not sufficient} for detecting CO.

We find no evidence for a threshold \HI\ column density above which the detection fraction suddenly increases, nor is there indication of a ``saturation'' \HI\ column density above which additional gas appears purely in molecular form.  In particular, there is a large range in $\log N_{\rm H}$ (from 21--21.8) over which CO is detected, and the lack of CO detections at lower $N_{\rm H}$ may be a reflection of our sensitivity limit.  As demonstrated below (\S\ref{sec:press}), however, the {\it azimuthally averaged} value of $N_{\rm H}$ is close to $10^{21}$ cm$^{-2}$ ($\approx$10 \Msol\ pc$^{-2}$) across a large range in radius, so one might easily conclude from the radial \HI\ profile that this represents a ``saturation'' value, even though locally much higher \HI\ column densities are observed.

The error bars shown in the detection fraction histograms are derived as follows.  First, the uncertainty in the number of pixels in a bin is assumed to arise from counting errors, so it is given by $\sqrt{N}$.  The uncertainty in the number of CO detections in a bin is estimated as the number of pixels with values between 3.5$\sigma$ and 4.5$\sigma$, where $\sigma$ is the RMS noise of the CO map.  Finally, the uncertainty in the detection fraction is given by standard error propagation assuming normally distributed errors.  The dashed histograms represent the range of detection fractions observed when separately analyzing the four different pixel samples we have available, since we only select a quarter of the original pixels for analysis.  The quoted errors appear to be consistent with these values, if not overly conservative.

The slanted line in Figure~\ref{fig:corrmom}, labeled $f_{\rm mol}$=0.5, indicates equality between atomic and molecular column densities (expressed in terms of $N_{\rm H}$).  Evidently the column density of molecular gas rarely exceeds the column density of atomic gas; the LMC appears to be an \HI-dominated galaxy.  (Although the upper limits on the left side of the figure formally allow high molecular fractions in this region of parameter space, we note that if CO emission actually existed at the 3$\sigma$ level for all non-detected pixels, the total CO flux would exceed the measured flux by a factor of 9).  We caution that if the CO-to-H$_2$ conversion factor is larger in the LMC than in the Galaxy, as expected from its low metallicity and dust content, the molecular column densities will be underestimated, and the equality line will shift downward (e.g., by 0.6 dex for a factor of 4 change in $N_{\rm H}$).  This would increase the number of molecule-dominated regions, though they would remain in the minority.  On the other hand, the \HI\ column density could also be underestimated if the \HI\ opacity is significant.  We return to a discussion of these uncertainties in \S\S\ref{disc:xfac}--\ref{disc:hiopac}.

\subsubsection{\HI\ Peak Brightness Temperature}

High peak \HI\ brightness temperatures also appear to be a requirement for CO detection, as indicated by Figure~\ref{fig:corrpk}(a).  A minimum cutoff level of $T_{\rm max}$=2.4 K has been applied, corresponding to 3$\sigma$ in the \HI\ cube.  Bright CO is detected almost exclusively in regions where $T_{\rm max}>20$ K, and the likelihood of detection increases monotonically with $T_{\rm max}$.  Once again, however, there is no indication of a threshold in $T_{\rm max}$ for CO detection, and a clear one-to-one correlation is lacking: many high-brightness \HI\ filaments are not detected in CO.  The Spearman rank correlation coefficient between $T_{\rm max}$(\HI) and $I_{\rm CO}$ is only $r_{\rm det}=0.15$; including upper limits yields a similar value ($r_{\rm surv}=0.14$), though with somewhat greater significance.

Although $T_{\rm max}$ will scale with $I_{\rm HI}$ for constant \HI\ line width, it can also highlight regions of significant \HI\ opacity: in the Galaxy, sight lines with $T_{\rm max} > 60$ K typically show \HI\ in absorption with $\tau \sim 1$ \citep{Braun:92}.  \citet{Braun:97} has argued that $T_{\rm max}$ images of nearby spiral galaxies serve as direct probes of the cool neutral medium (CNM), the \HI\ phase most similar in temperature and density to molecular clouds.  To check if beam dilution was responsible for the low values of $T_{\rm max}$ observed towards some CO-detected regions, we extracted the 38 CO-detected pixels with $T_{\rm max} < 30$ K (just 5\% of all CO detections) and measured their \HI\ peak brightness temperatures at the full 1\arcmin\ resolution of the ATCA data.  The average value of $T_{\rm max}$ at 1\arcmin\ resolution is 31~K, compared to 23~K at 2\farcm6 resolution.  Thus, a small but significant number of regions appear to show detectable CO without high brightness \HI, although the general conclusion remains that high values of $T_{\rm max}$ are necessary for detecting CO.

\begin{figure*}
\includegraphics[width=8cm]{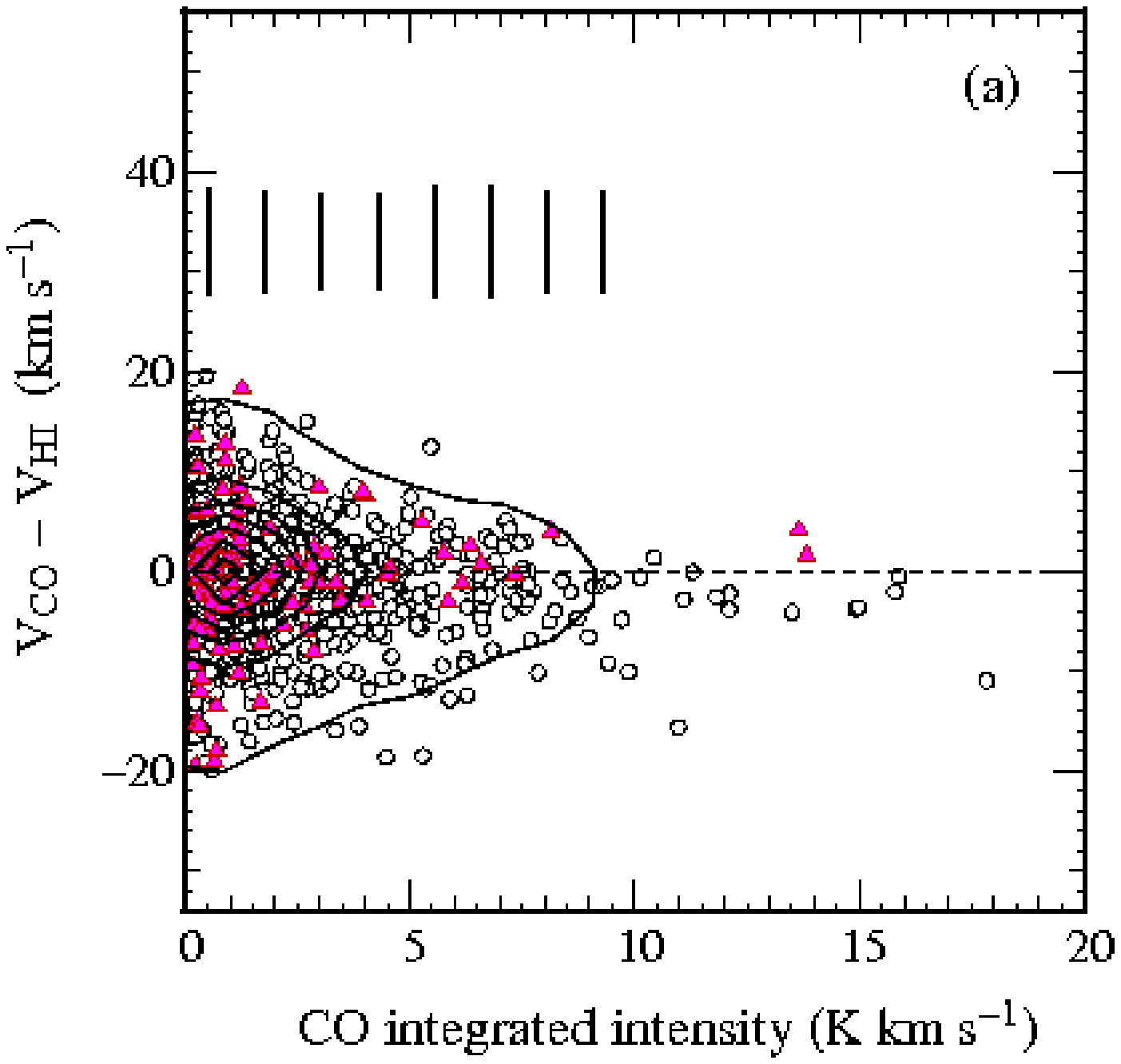}\hfill
\includegraphics[width=8cm]{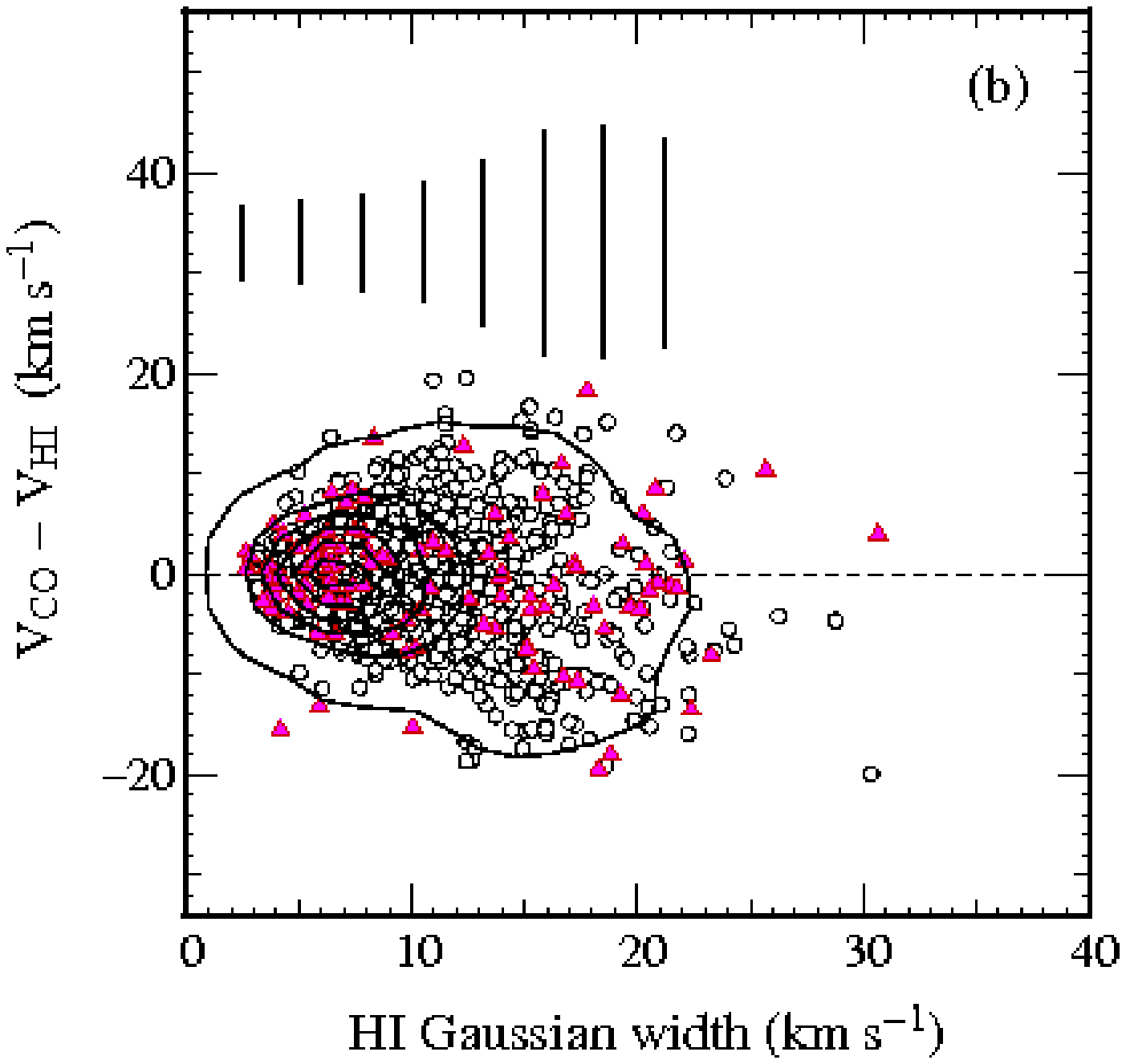}
\caption{
Scatter plots of the CO-\HI\ velocity difference as a function of (a) CO intensity and (b) width of the associated Gaussian \HI\ component.  Open circles represent pixels where the dominant (highest peak $T_b$) \HI\ component is associated in velocity with the CO, while filled triangles represent pixels where the secondary component is associated.  Contours indicate the density of points (including both components), and the lengths of vertical bars above the plotted points indicate the FWHM of the distribution.
\label{fig:vdiff}}
\end{figure*}

\begin{figure*}
\includegraphics[width=8cm,bb=32 156 575 680]{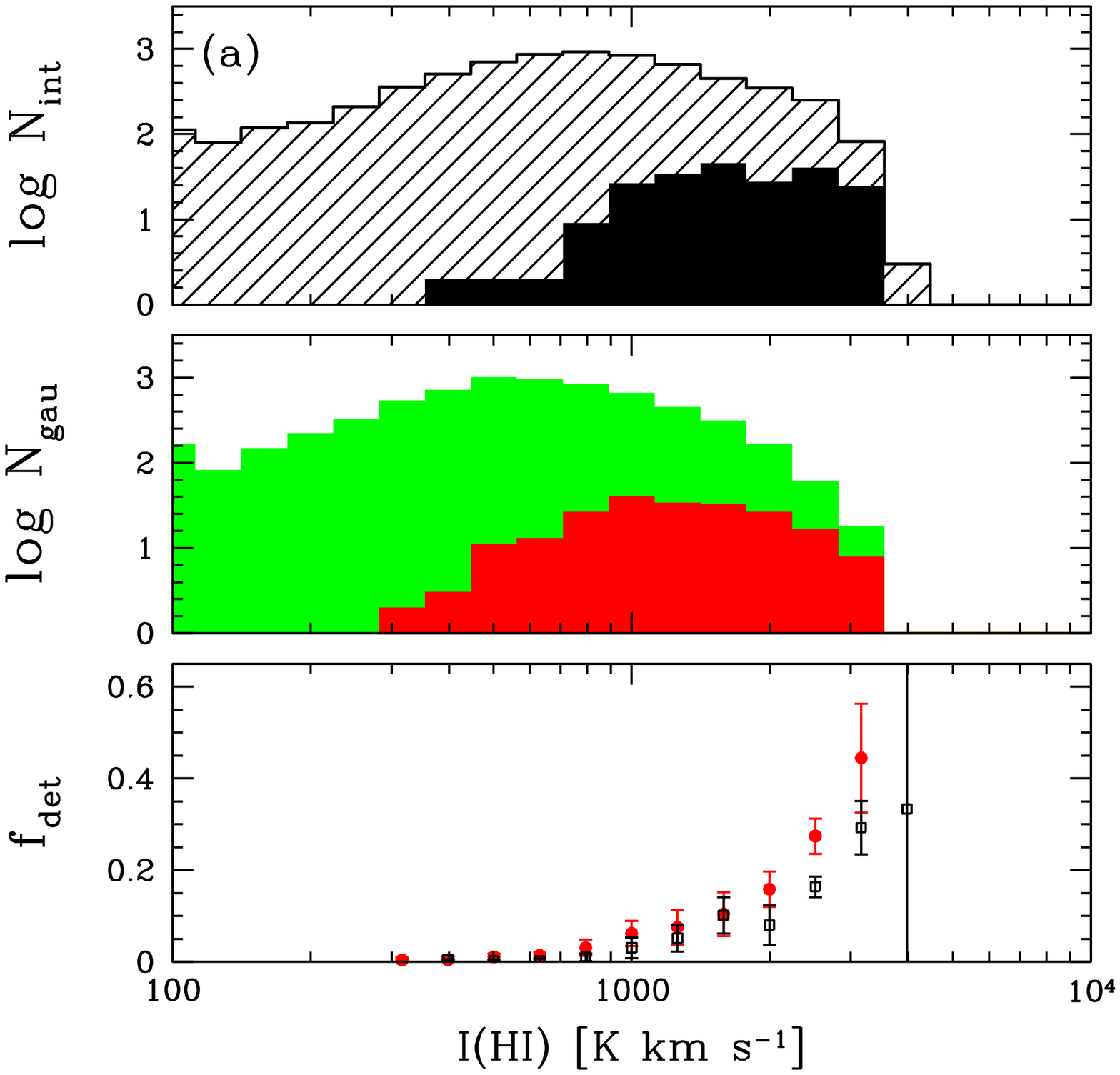}\hfill
\includegraphics[width=8cm,bb=32 156 575 680]{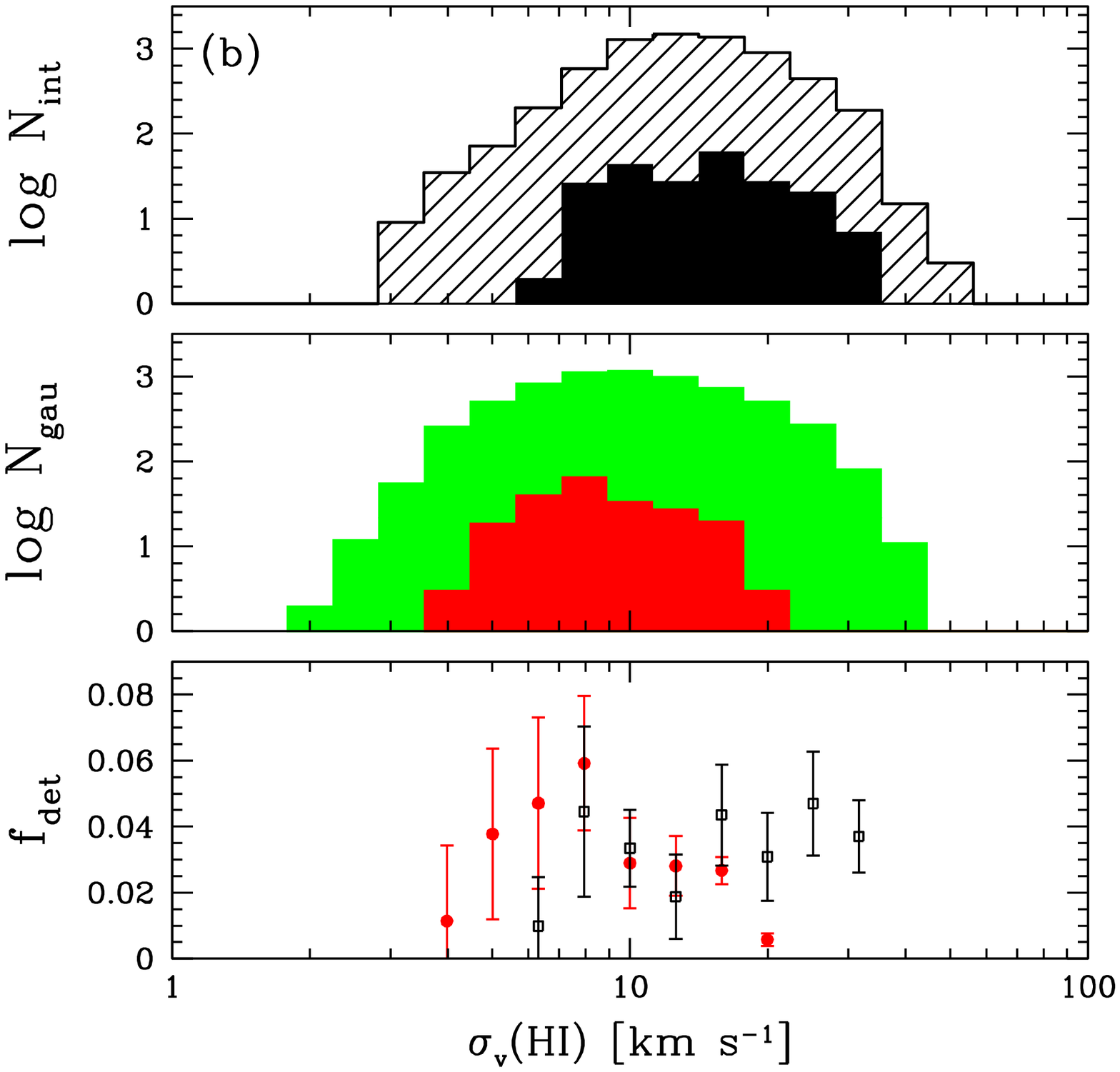}
\caption{
Histograms of (a) \HI\ intensity and (b) \HI\ velocity dispersion for the entire region observed in CO.  The top sub-panels show the distribution of all pixels using diagonal shading and the distribution of CO-detected pixels using black shading.  The middle sub-panels show the same two distributions (as green and red shaded histograms respectively) but with the \HI\ restricted to the dominant component in a two-Gaussian decomposition.  The bottom sub-panels show, on a linear scale, the fraction of pixels detected in CO, with \HI\ restricted to the dominant Gaussian component shown as solid red circles.
\label{fig:corrgint2}}
\end{figure*}

\begin{figure*}
\begin{center}
\includegraphics[width=7.5cm]{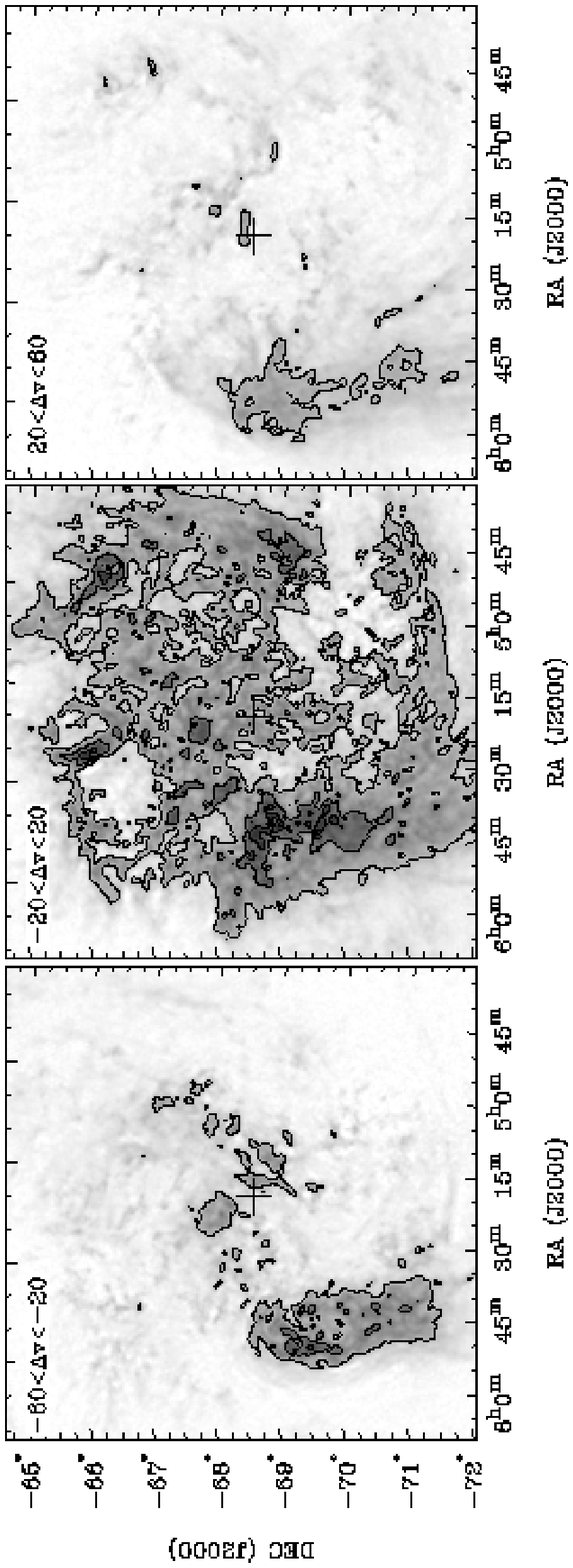}\hspace{1.5cm}
\includegraphics[width=7.5cm]{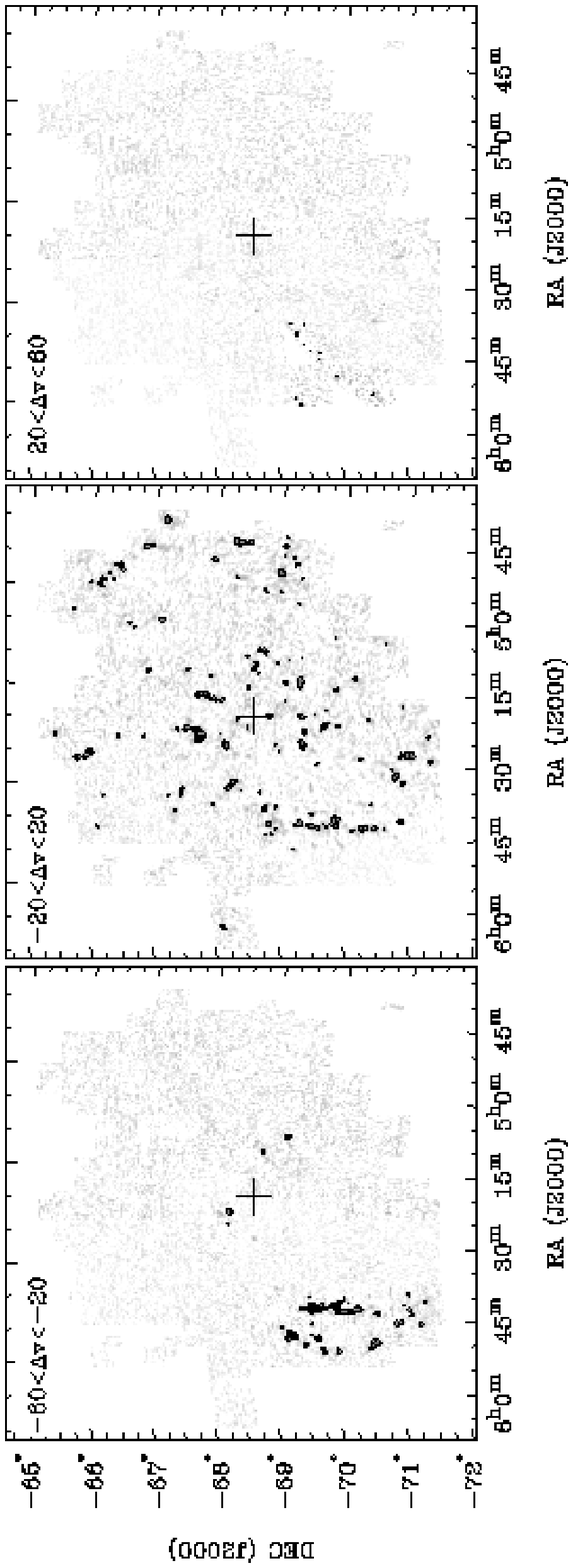}
\end{center}
\caption{
\HI\ (top) and CO (bottom) channel maps integrated over three velocity ranges given with respect to a circular rotation model.  Contour levels are 400$n$ K \kms\ for \HI\ and 2.4$n$ K \kms\ for CO, where $n$=1,3,5,7.  A cross indicates the adopted rotation center.
\label{fig:vellmc}}
\end{figure*}

\subsubsection{\HI\ Velocity Dispersion}\label{sec:hidisp}

In contrast to the integrated intensity and peak brightness, the \HI\ line width, as estimated by the 2nd moment of the line profile (excluding pixels below a 4$\sigma$ threshold in $I_{\rm HI}$, and rejecting linewidths $<$3 \kms), shows no relationship with CO [Figure~\ref{fig:corrpk}(b)].  Thus, \HI\ velocity dispersion is a poor predictor of whether CO emission will be detected.  One interpretation of this result is that colliding \HI\ flows are unlikely to be the direct precursors to GMCs.  Alternatively, the 2nd moment of the \HI\ line profile may not reflect the local \HI\ velocity dispersion at any position in space, due to blending of multiple gas components along the line of sight.  This possibility is addressed in \S\ref{sec:gaufit}, where we deduce a weak {\it anti-correlation} between $\sigma_v({\rm HI})$ and CO detectability following a Gaussian decomposition of the line profile.  Our results indicate that if GMCs form from colliding \HI\ flows, then much of the turbulent kinetic energy dissipates before CO becomes easily visible.

\subsection{Effect of Varying Spatial Resolution}\label{sec:smores}

To determine whether the poor point-by-point correlations between CO and \HI\ found in \S\ref{results:props} result from a strong global correlation breaking down on small spatial scales, we spatially binned the original 2\arcmin\ pixels into 8\arcmin\ square and then 16\arcmin\ square pixels and repeated the correlation analysis.  Some improvement in the correlation is to be expected, simply by virtue of blending distinct emission peaks so as to emphasize the large-scale galactic structure.  We indeed find larger correlation coefficients ($r_{\rm surv}$=0.25 and 0.37 respectively), but caution that the range of values spanned by the permutation tests also increases.  We also see a modest increase in the fraction of the ``total'' (integrated over the entire datacube) CO flux recovered in the CO-detected pixels, from 54\% to 67\% (8\arcmin\ pixels) and 74\% (16\arcmin\ pixels), although such measurements are sensitive to any errors in the spectral baseline.  Finally, we note an increase in the fraction of CO-detected pixels (Figure~\ref{fig:bin48}).  This appears to reflect the basic tendency for detection within any of the pixels contributing to a bin to result in the larger, averaged pixel being considered a detection.  In fact, of the 8\arcmin\ pixels observed in both CO and \HI, only 0.4\% were classified as detections when none of the 16 constituent 2\arcmin\ pixels were so classified (these are shown as solid green circles in Figure~\ref{fig:bin48}).  Thus, the increased detection fraction after binning is not due to combining pixels that lie just below our detection limits.

As indicated by Figure~\ref{fig:bin48}, measuring the correlation on a coarser grid still leads to a large range in $I_{\rm CO}$ at a given $I_{\rm HI}$, even as the overall CO detection fraction increases.  In particular, there is little evidence for a change in the tightness of the correlation above or below a particular averaging scale.  This confirms the visual impression from Fig.~\ref{fig:maps} that the differences between the \HI\ and CO distributions exist on large as well as small scales.  Since most of additional flux that is recovered by spatial averaging appears to be located close to CO peaks already detected at high resolution, any additional flux tends to correlate with the \HI\ emission in a similar way.

\subsection{Association in Velocity Space}\label{sec:gaufit}

Our approach of integrating in velocity takes no account of whether \HI\ and CO are actually coincident in velocity space.  In Paper II, we will undertake a full 3-D comparison of the data cubes.  However, we can test whether velocity integration has strongly skewed our results by performing an automated Gaussian decomposition of the \HI\ cube using the GAUFIT task in the MIRIAD software package \citep{Sault:95}.  While such an analysis cannot guarantee that we are able to compare atomic and molecular gas at the same line-of-sight position, it makes it less likely that a correlation (or lack thereof) is solely due to projection effects.  We fit up to two Gaussian components at each sky position, and if CO is detected at the same position, the \HI\ component closer to the CO velocity (if within 20 \kms) is taken to be {\it associated} with CO.\@  The component with the highest amplitude (peak $T_b$) is taken to be the {\it dominant} \HI\ component.  For about 1/6 of the pixels we find a significant improvement in $\chi^2$ when going from single to double Gaussian fits, especially in the central and southeastern parts of the LMC, where two widely separated \HI\ components have been noted \citep{Luks:92}.

For purposes of comparison with the \HI\ Gaussian fits, we assume the CO profiles are single-peaked, which is a good assumption for the vast majority of spectra, for which fitting an additional component has little impact on $\chi^2$.  In order to minimize noise contamination in determining the radial velocity at the peak of each CO spectrum, we applied a signal detection algorithm adapted from the IDL-based CPROPS package of \citet{Rosolowsky:06}.  A masked datacube is generated by starting at the 3$\sigma$ contour (required to encompass at least two consecutive channels and three adjacent pixels) and expanding it to include adjacent pixels in position and velocity above a 2$\sigma$ level.  The CO velocity is then taken as the velocity at the peak of each spectrum within the masked datacube.  Of the 479 CO spectra in the full-resolution map that at least partially survive the masking process, we find that the vast majority (95\%) associate closest in velocity with the dominant (or sole) \HI\ component, with only 4\% associated with the lower amplitude component.  Only 1\% of the CO detections were unassociated with \HI\ within 20 \kms; these spectra were usually of low signal-to-noise or found near 30 Doradus, where the strong 1.4 GHz continuum leads to strongly inverted \HI\ profiles that cannot be fit by the Gaussian decomposition.

Although by definition the velocity difference between associated CO and \HI\ components is less than 20 \kms, in most cases the difference is considerably smaller (Figure~\ref{fig:vdiff}).  In fact, for \HI\ profiles best fit by two Gaussians, a significant fraction (19\%) of the CO detections lie within 20 \kms\ of both \HI\ Gaussian components.  No clear relation between the velocity difference and the CO intensity is apparent [Fig.~\ref{fig:vdiff}(a)], with the distribution of velocity differences exhibiting a nearly constant width of $\sim$10 \kms, independent of the CO intensity.  Not surprisingly, many of the largest velocity differences occur when the associated \HI\ component has a large line width, in which case the velocity centroid may be less clearly defined or several unresolved \HI\ components may be superposed along the line of sight [Fig.~\ref{fig:vdiff}(b)].

How are the detection histograms in Figures~\ref{fig:corrmom} and \ref{fig:corrpk} affected by the Gaussian decomposition?  In the upper panels of Figure~\ref{fig:corrgint2} we show the distributions of \HI\ intensities and velocity dispersions over the full region observed by NANTEN and over just the CO-detected pixels.  The middle panels show the equivalent distributions after isolating only the dominant Gaussian component (here the velocity dispersion is measured as the Gaussian width, which is equivalent to the 2nd moment).  As expected, the Gaussian decomposition shifts both the \HI\ intensity and velocity dispersion to lower values.  In the bottom panels, a marginal increase in the CO detection rate at the highest \HI\ intensities is apparent following decomposition, underscoring the fact that CO associates with the dominant \HI\ component.  On the other hand, the CO detection fraction drops to zero for high \HI\ velocity dispersions, even in bins which remain highly populated.  Thus, the Gaussian decomposition appears to reveal an {\it anti-correlation} between $\sigma_v$ and CO detectability, although in all cases the CO detection fraction is low ($\lesssim$6\%).  The shift in CO detections toward lower $\sigma_v$ suggests that many of the detections are in regions where the total line profile is artificially broadened (as measured by the 2nd moment) due to the presence of multiple \HI\ components.

\begin{figure}
\includegraphics[height=9cm]{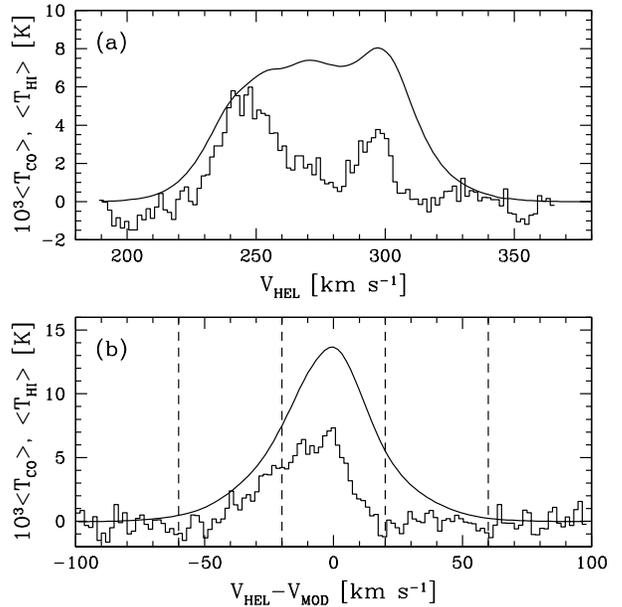}
\caption{
(a) Mean CO (histogram) and \HI\ (curve) brightness temperature as a function of heliocentric velocity, calculated over the area of sky shown in Fig.~\ref{fig:vellmc}.  The CO data have been scaled up by a factor of $10^3$.  (b) Same as (a), but after shifting the velocity scale of each spectrum to be centered on a circular rotation model.  The dashed lines represent the boundaries of integration used for Fig.~\ref{fig:vellmc}.
\label{fig:imspec}}
\end{figure}

To emphasize that CO emission primarily occurs near the velocity of the dominant \HI\ component, regardless of how regular that component appears in velocity space, we have fit a circular rotation model to the \HI\ line-of-sight velocity field, estimated using the 1st moment of the line profile.  Given the large angular size of the galaxy, we first correct for the transverse motion of the LMC, following the prescription of \citet{vdMarel:02}, using the proper motion vector derived by \citet{Kalliv:06} using {\it Hubble Space Telescope} imaging.  We assume a disk inclination of 35\arcdeg\ and, using the ROTCUR task in the GIPSY software package, derive a mean kinematic position angle of 340\arcdeg\ and a kinematic center of 5$^{\rm h}$19\fm5, $-$68\arcdeg 53\arcmin\ (J2000), about 14\arcmin\ from that determined by \citet{Kim:98} (this difference, while large, is within the formal errors of the fit).  
Shifting the \HI\ spectra to the mean velocity of the circular rotation model yields a prominent low-velocity excess in the central and southeastern regions (Figure~\ref{fig:vellmc}), which can be identified as the ``L'' component of \citet{Luks:92}.  In addition, a (weaker) high-velocity component is also apparent, suggesting a bifurcation in the gas layer at the eastern edge of the galaxy, as discussed by \citet{Nidever:08}.  Shifting the CO spectra according to the same model, we find that the CO is {\it not} exclusively associated with the main disk component: at the eastern edge of the galaxy, it is associated primarily with the L component, as was previously noted by \citet{Mizuno:01}.  The CO excess at low velocities is also apparent in spatially averaged spectra, both before and after subtraction of the circular rotation model (Figure~\ref{fig:imspec}).

\begin{figure*}
\begin{center}
\includegraphics[height=8cm,angle=0]{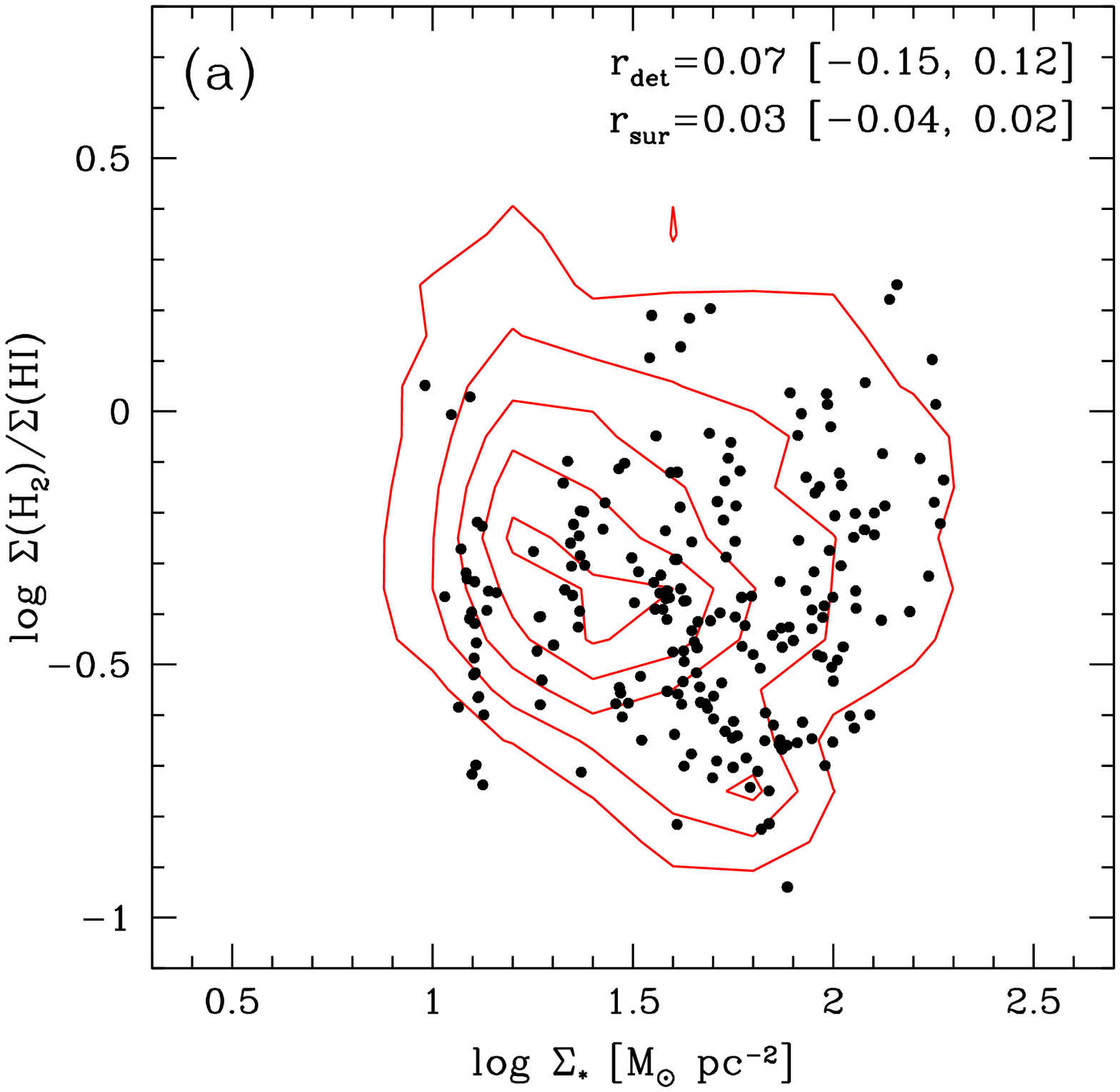}
\includegraphics[height=8cm,angle=0]{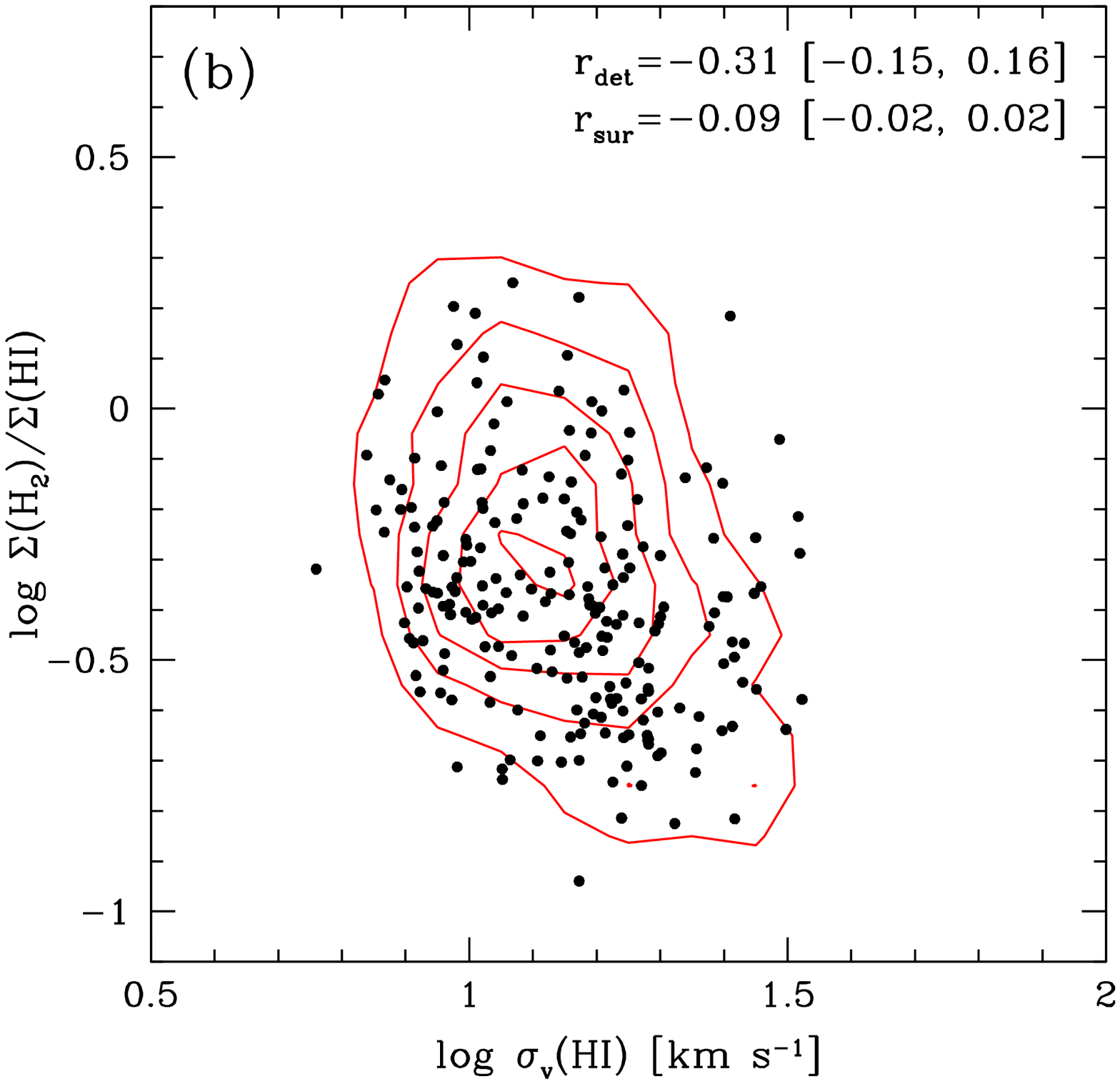}\\
\includegraphics[height=8cm,angle=0]{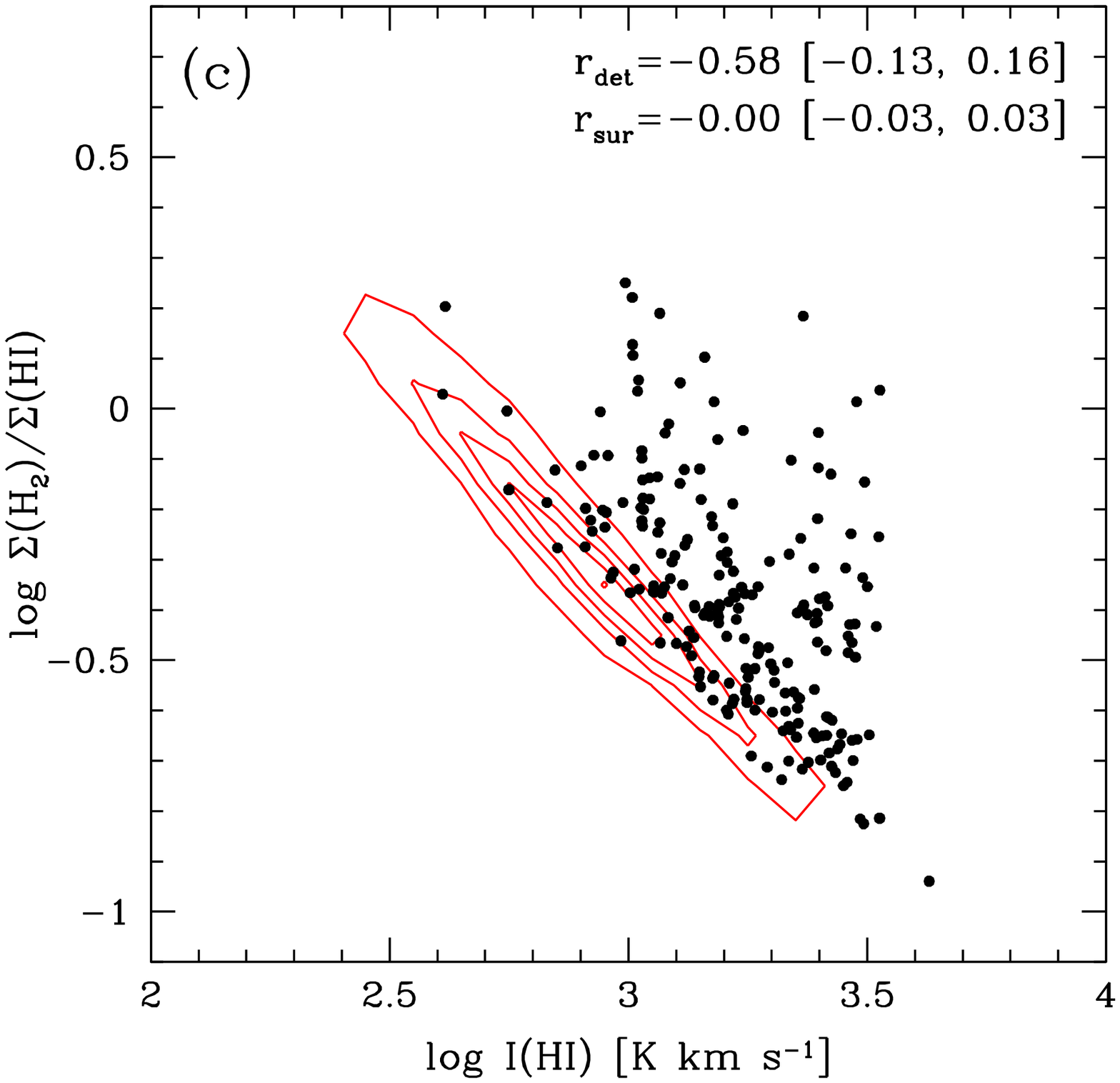}
\includegraphics[height=8cm,angle=0]{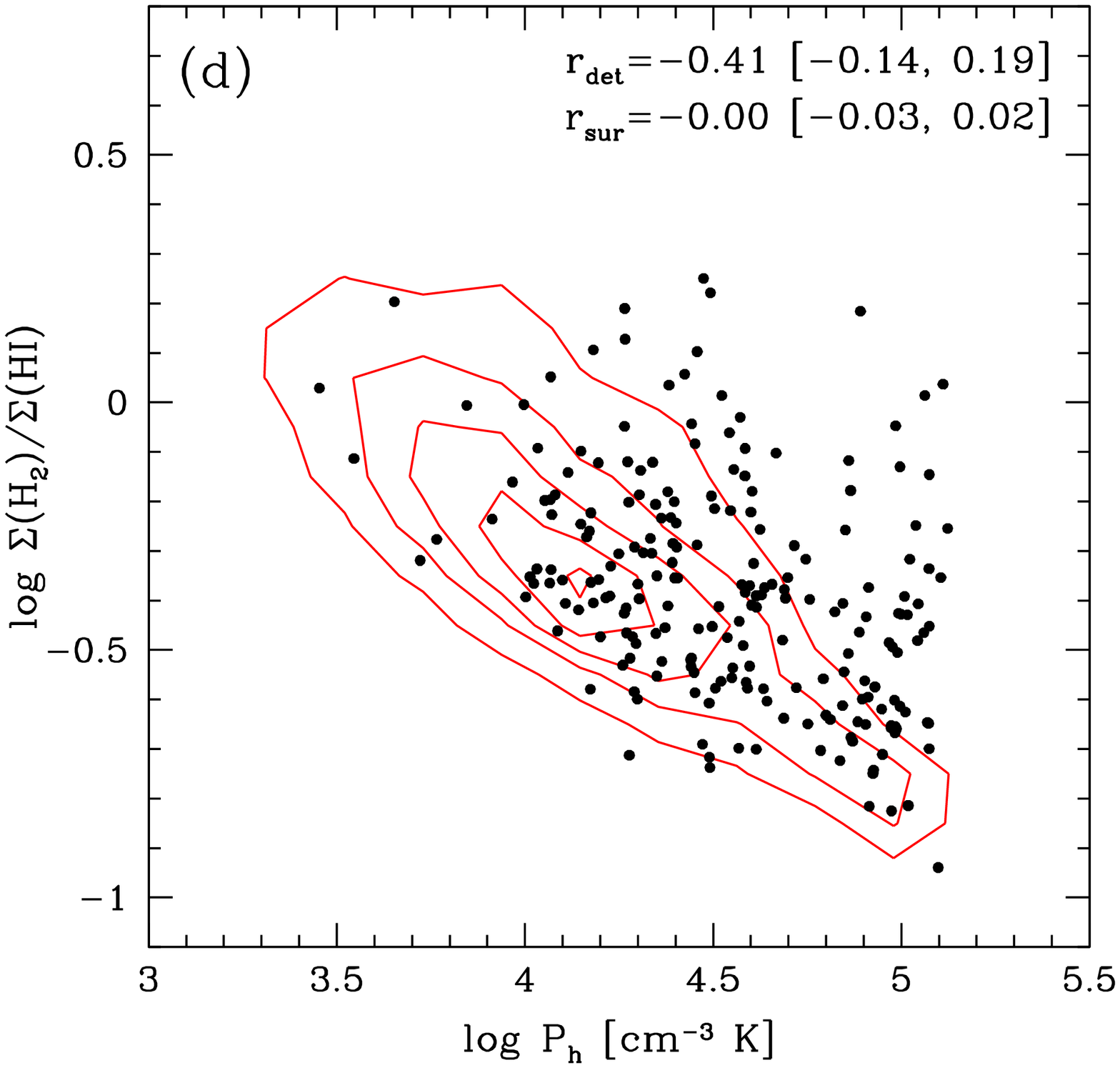}\\
\end{center}
\caption{
Ratio of molecular to atomic gas surface density as a function of (a) stellar surface density; (b) \HI\ velocity dispersion; (c) \HI\ integrated intensity; (d) average midplane hydrostatic pressure, calculated as described in the text.  The CO detections are shown as solid circles.  Contours, spaced between 15\% and 95\% of the peak in increments of 20\%, indicate the density of points for {\it upper limits only}, corresponding to CO non-detections.  Correlation coefficients are calculated as for Figures~\ref{fig:corrmom}--\ref{fig:bin48}.
\label{fig:rmolstar}}
\end{figure*}

\subsection{Molecular to Atomic Ratio}\label{sec:press}

\citet{Wong:02}, \citet{Blitz:04,Blitz:06} and \citet{Leroy:08} have examined \HI\ and CO in nearby spiral galaxies and found that the ratio of molecular to atomic gas surface density correlates strongly with the hydrostatic pressure at the disk midplane, $P_h$, estimated for a two-component disk of gas and stars.  As given by \citet{Elmegreen:89}, $P_h$ can be estimated from the observed mass surface densities of gas and stars ($\Sigma_{\rm gas}$ and $\Sigma_*$) as follows:
\begin{equation}
P_h = \frac{\pi G}{2} \Sigma_{\rm gas} \left( \Sigma_{\rm gas} + \frac{\sigma_g}{\sigma_*} \Sigma_* \right) \;,
\label{eqn:hydro}
\end{equation}
where the term in parentheses is an estimate of the total dynamical mass within the gas layer, and $\sigma_g$ and $\sigma_*$ are the velocity dispersions of gas and stars, respectively.  

\begin{figure}
\includegraphics[height=9cm]{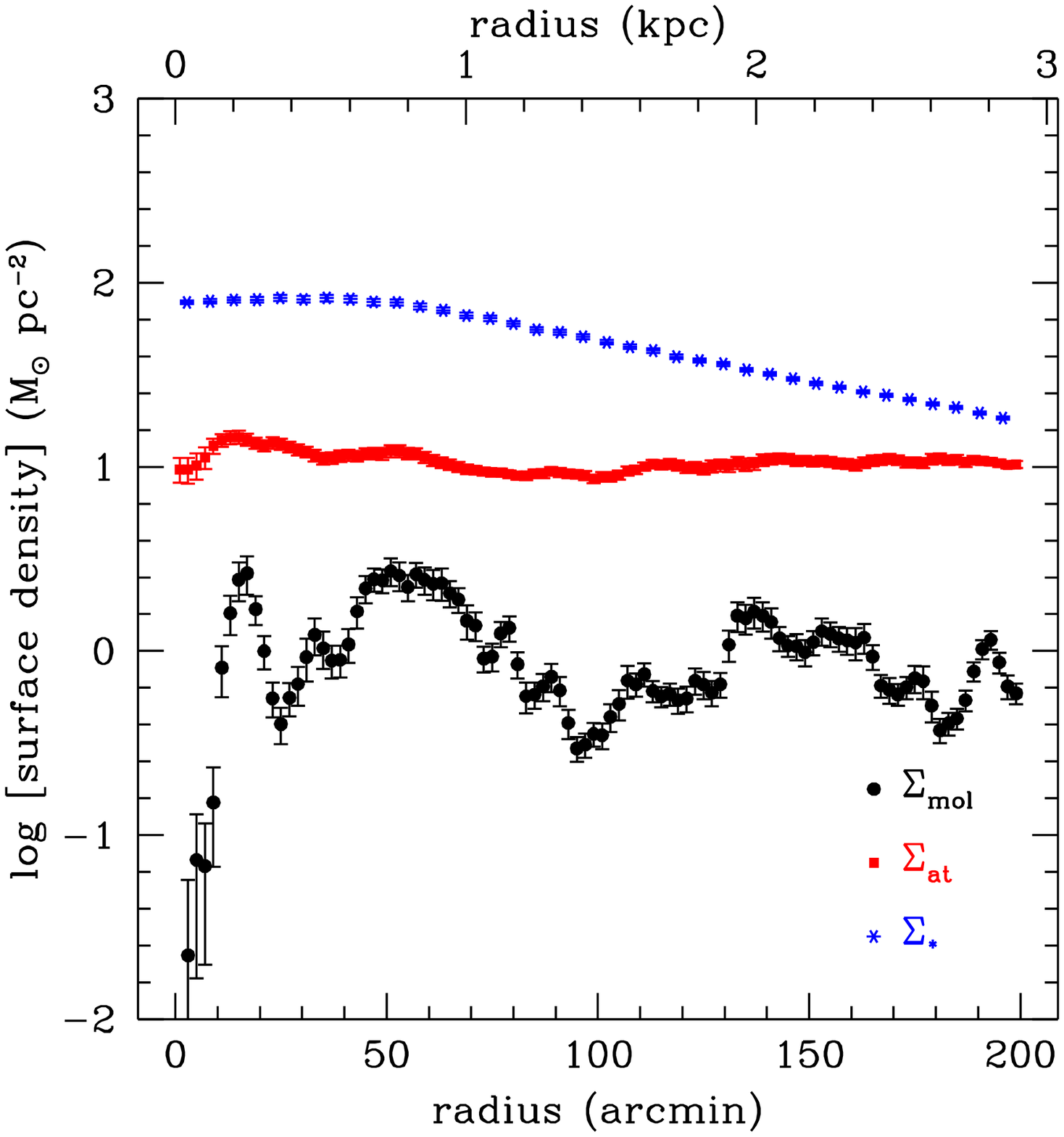}
\caption{
Azimuthally averaged radial profiles of stellar, \HI\ and H$_2$ mass surface density for the LMC.  A Galactic value for the CO-to-H$_2$ conversion factor has been assumed.  The adopted center is 5$^{\rm h}$19\fm5, $-$68\arcdeg 53\arcmin\ (J2000) and the adopted inclination and major axis position angle are 35\arcdeg\ and 340\arcdeg\ respectively.
\label{fig:radprof}}
\end{figure}

To investigate the possible role of pressure in determining the H$_2$/\HI\ ratio $R_{\rm mol}$ in the LMC, we first consider two quantities contributing to $P_h$ that can be determined independently of the gas surface density.  In Figure~\ref{fig:rmolstar}(a) the inferred molecular to atomic surface density ratio, calculated at pixels where $I_{\rm HI} > 110$ K \kms, is shown as a function of the stellar surface density $\Sigma_*$, which dominates the gravitational potential in the inner disk.  CO detections are shown as solid points, whereas the much larger number of CO upper limits have been binned and contoured.  Thus, the contours should be thought of as the upper envelope of the CO non-detections.  We have used for $\Sigma_*$ the stellar surface density image of \citet{Yang:07}, which is based on number counts of red giant and asymptotic giant branch stars as identified by 2MASS colors, and normalized to a total stellar mass of $2 \times 10^9$ \Msol.  No correlation between $R_{\rm mol}$ and $\Sigma_*$ is apparent, nor is there any clear separation between the CO detections and non-detections.  Similarly, as shown in Figure~\ref{fig:rmolstar}(b), only a weak anti-correlation is found when comparing $R_{\rm mol}$ with the \HI\ velocity dispersion, $\sigma_v$(\HI), calculated from the 2nd moment of the line profile.  Using the Gaussian dispersion of the dominant \HI\ component for $\sigma_v$ does not significantly alter this result.

In Figure~\ref{fig:rmolstar}(c) we correlate $R_{\rm mol}$ against $I_{\rm HI}$, on a pixel-by-pixel basis, reproducing Figure~\ref{fig:corrmom} but now with a substantial ``tilt'' that results from dividing by the abscissa.  The detections and non-detections are clearly separated, but no clear relationship is evident when taking the upper limits into account.  Evidently, the apparent anti-correlation seen in the detected points can be attributed to the difficulty in measuring $R_{\rm mol}$ when $I_{\rm HI}$ is weak.  Finally in Figure~\ref{fig:rmolstar}(d) we correlate $R_{\rm mol}$ against $P_h$, with only \HI\ considered in calculating $\Sigma_{\rm gas}$.  This should be a reasonable approximation for the LMC because the contribution of H$_2$ to the total gas mass is relatively small unless an extremely large CO-to-H$_2$ conversion factor is assumed.  We plot radial mass surface density profiles for stars, atomic gas, and molecular gas in Figure~\ref{fig:radprof} to verify this; the inferred H$_2$ surface density is generally a factor of 5--10 less than the \HI\ surface density.  We assume a constant velocity dispersion of 9 \kms\ for the \HI, based on the average dispersion of the dominant Gaussian component, and 20 \kms\ for the stars \citep{vdMarel:02}.  Overall the results are similar to those shown in 
Figure~\ref{fig:rmolstar}(c), with again an apparent anti-correlation likely driven by a strong tilt in the distribution of upper limits.  We therefore find no clear evidence for a relationship between $R_{\rm mol}$ and either $\Sigma_{\rm gas}$ or $P_h$.

An obvious limitation of this analysis is the preponderance of high upper limits on $R_{\rm mol}$.  This reflects the comparatively poor sensitivity of the CO data, which precludes useful constraints on the CO/\HI\ ratio where \HI\ emission is weak.  By contrast, the adopted \HI\ brightness cutoff when calculating $R_{\rm mol}$, corresponding to a column density of $N_{\rm H}$=$2 \times 10^{20}$ cm$^{-2}$, only excludes 2\% of the pixels in the region observed in CO.\@  One could increase the signal-to-noise ratio on $R_{\rm mol}$ by azimuthally averaging $\Sigma_{\rm gas}$, as has been done for spiral galaxies by \citet{Wong:02}, but using the radial profiles shown in Figure~\ref{fig:radprof} fails to reveal any meaningful trends between $R_{\rm mol}$ and $P_h$, given that the large fluctuations in the radial CO profile are not reflected in either the \HI\ or the stellar profile.  In \S\ref{disc:press} we average $R_{\rm mol}$ in bins of increasing $P_h$ to provide an alternative comparison to the spiral galaxy results.

\begin{figure*}
\includegraphics[width=\textwidth]{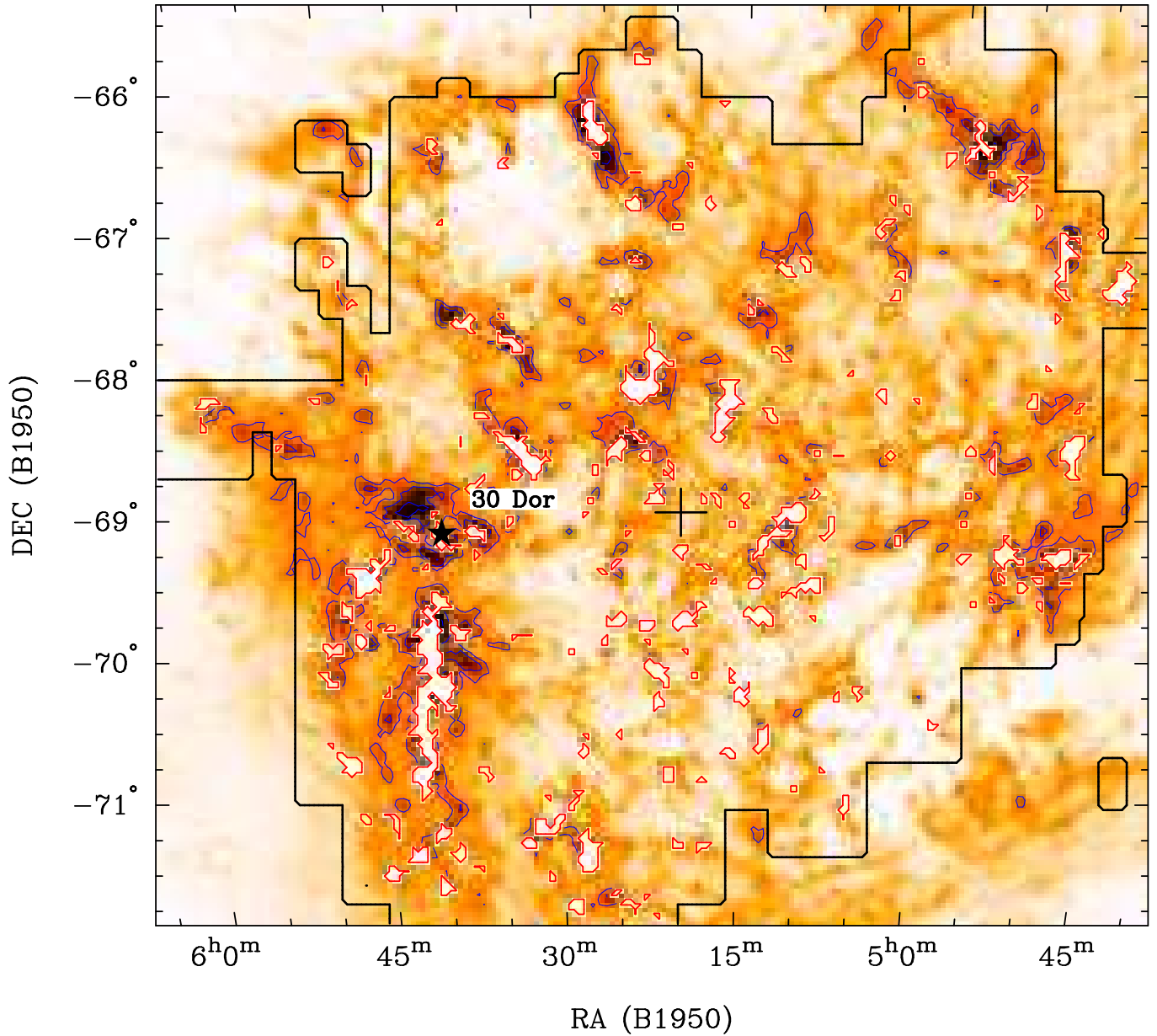}
\caption{
Image of peak \HI\ brightness temperature, with regions where CO is detected (as determined by the CPROPS mask described in \S\ref{sec:gaufit}) excluded (shown as white regions).  Contours are spaced by 20~K starting at 60~K.  The location of the 30 Doradus star-forming region is indicated by a star, and the kinematic center is indicated by a cross.
\label{fig:hinoco}}
\end{figure*}

\section{Discussion}\label{disc}

\subsection{Interpreting the CO-\HI\ Correlation}\label{disc:cohi}

We have found that while \HI\ appears to be a necessary condition for the appearance of CO emission, the intensities of the two tracers are poorly correlated across the LMC.  Given the expected core-envelope structure discussed in \S\ref{sec:intro}, this is somewhat surprising.  In the absence of significant overlap between clouds, an increased number of molecular cloud cores should be accompanied by an increased mass in \HI\ envelopes.  Conversely, high \HI\ column densities should promote H$_2$ formation because more gas can in principle be shielded from external UV radiation.  The absence of a strong correlation leads to three general classes of interpretation, each of which can be linked with one or more physical mechanisms.

The first possibility is that the scatter in the correlation results from a partial depletion of \HI\ as it is converted to H$_2$.  This could generate a large observed range in the ratio of CO to \HI\ intensity, depending on how far the conversion proceeds.  Recently, \citet{Ott:08} have reported evidence of \HI\ to H$_2$ conversion, in the form of offsets between CO and \HI\ peaks, for molecular complexes south of 30 Doradus mapped at 45\arcsec\ resolution.  However, the larger scales probed by our analysis (by a factor of 4 in each linear dimension) should dilute the effects of such localized conversions, because the \HI\ appears to dominate the gas mass both locally (Figure~\ref{fig:corrmom}) and on large scales (Figure~\ref{fig:radprof}).  Formation of H$_2$ therefore seems unlikely to have had a major effect on the \HI\ intensity at the scales we are probing.

The second possibility is that our assumption that CO and \HI\ trace molecular and atomic column density faithfully is flawed.  For instance, the dense gas may be incompletely sampled by CO emission, due to the time required for CO to become visible or variations in the degree of shielding against dissociating FUV radiation, or may be underluminous in the \HI\ line due to high optical depth.  We discuss these possibilities in greater detail in \S\S\ref{disc:xfac} and \ref{disc:hiopac} below.

The third possibility is that volume density and column density are decoupled: large \HI\ column densities do not necessarily imply the large volume densities characteristic of molecular clouds.  This decoupling would arise naturally from a change in the scale height of the gas layer---due to fluctuations in the amplitude of turbulent motions or a variation in the gravitational potential---or may reflect differences in density contrast which are smoothed over when averaging over $\sim$40 pc scales.  We discuss these possibilities in greater detail in \S\S\ref{disc:hz} and \ref{disc:warmhi} below.

\subsubsection{Variations in the CO-to-H$_2$ Ratio}\label{disc:xfac}

The use of CO as a tracer of H$_2$ is usually justified with the following argument: while the CO(1--0) line is optically thick, the integrated line emission arises from an ensemble of virialized clouds that can be characterized by a common value of excitation temperature and density, and that do not shadow each other in position and velocity space \citep{Maloney:88,Papadop:02}.  Even adopting these assumptions, however, it is unclear how reliably CO intensities can be translated into H$_2$ column densities for {\it individual} clouds, which would be expected to show variations in temperature, density, and degree of virialization.  The role of optical depth variations is also unclear.  For clouds which are less optically thick, the observed emission may come from deeper, warmer layers, actually increasing the CO brightness.  While we lack the data to address these issues properly, we note that multi-transition studies of LMC clouds have generally found line intensity ratios that are consistent with $^{12}$CO having significant opacity \citep[e.g.,][]{Johansson:94,Israel:03}. \citet{Fukui:08} find a strong correlation between CO luminosity and virial mass for the NANTEN cloud sample, and thus derive a CO-to-H$_2$ conversion factor of $7 \pm 2 \times 10^{20}$ cm$^{-2}$ (K \kms)$^{-1}$, roughly twice the value obtained from application of the virial theorem to Galactic clouds \citep{Solomon:87}, and more than three times the value obtained from Galactic $\gamma$-ray studies \citep{Strong:96}.

A particular concern with regard to low-metallicity systems such as the LMC is that the CO-emitting region of a molecular cloud could become significantly smaller than the cloud itself, despite the high opacity of the CO line, as a result of CO photodissociation.  Since dissociation of H$_2$ occurs primarily at specific UV wavelengths (the Lyman-Werner bands), H$_2$ can shield itself from photodissociation at sufficiently high column densities ($\gtrsim$$10^{15}$ cm$^{-2}$).  Much higher equivalent column densities are required for CO self-shielding to be important, however, due to its much lower abundance, making shielding by dust important for the survival of CO.  Calculations therefore suggest that CO may not be the primary reservoir of carbon in some molecular regions \citep{vDB:88}.  Especially in regions permeated by strong UV radiation fields, H$_2$ may be present without abundant CO, weakening any correlation between \HI\ and CO emission that might otherwise exist.  \citet{Maloney:88} and \citet{Pak:98} have presented models showing how the CO-emitting parts of molecular clouds can be embedded in much larger H$_2$ envelopes.  The importance of this effect is a strong function of the dust abundance (and thus metallicity) as well as radiation field.  

An examination of the location of bright \HI\ not associated with CO emission (Figure~\ref{fig:hinoco}) lends some support to this interpretation.  Contours delineating regions with $T_{\rm max}$(\HI)$>$60 K generally intersect or enclose CO-detected pixels; this implies that bright \HI\ emission is generally found adjacent to, if not coincident with, CO emission.  Thus, many regions of strong \HI\ emission may harbor H$_2$ clouds but be insufficiently shielded from dissociating FUV radiation to permit survival of CO.  The strong concentration of \HI\ near 30 Doradus may be an example of this effect.  Interestingly, the largest high-$T_{\rm max}$ complexes appear to be in the outer regions of the galaxy.  Another contributing factor may be differences in the formation timescales of H$_2$ and CO, which can create variations in the CO/H$_2$ ratio across a cloud.  Models of molecular cloud formation behind shock waves indicate timescales of $\sim$10$^7$ yr for CO to become abundant, several times longer than the formation timescale for H$_2$ \citep{Bergin:04}.  Again, this is a result of the more stringent shielding requirements for CO formation.  However, this effect ought to be suppressed when averaging over areas much larger than an individual cloud ($\sim$10 pc), as we have done in this study.

Observations of H$_2$ in absorption provide the most direct constraints on the amount of molecular material that is missed by CO surveys.  Unfortunately, since such observations are performed at ultraviolet wavelengths, they are limited by the number of suitable background sources and a bias towards lines of sight with relatively low extinction.  Analysis of the existing {\it FUSE} data (Welty et~al., in preparation; see also \citealt{Tumlinson:02}) yields relatively low H$_2$ column densities ($N_{\rm H} \lesssim 10^{21}$ cm$^{-2}$) compared to those traceable by CO emission, and thus the low CO(1--0) detection rates found toward those sight lines are to be expected.  Although there is always ambiguity in comparing emission and absorption measurements because they generally probe different volumes and different line-of-sight depths, deeper CO integrations towards these positions would be worthwhile.  In addition, observations of H$_2$ (or suitable proxies) in absorption towards more highly reddened sight-lines near CO clouds would allow a more definitive comparison between absorption and emission studies.

Recently, indirect evidence for an H$_2$ component outside of CO-emitting regions has been presented by \citet{Leroy:07} for the SMC and \citet{Bernard:08} for the LMC based on far-infrared measurements with {\it Spitzer}, which would otherwise imply a discrepancy in dust abundance between low and high column density regions.  On the other hand, UV absorption line measurements \citep[and references therein]{Cartledge:05} have inferred just such a discrepancy by comparing $E(B-V)/N$(\HI) and $E(B-V)/N({\rm H}_2)$ toward multiple UV sight lines.  Indeed, one might expect a higher dust abundance in H$_2$-dominated regions because of the important role of dust grains in H$_2$ formation.  Thus, it remains unclear to what extent an unseen H$_2$ component is required by observational data.

\subsubsection{Variations in \HI\ Optical Depth}\label{disc:hiopac}

For an optically thick line profile, the integrated intensity will underestimate the column density.  In principle, high opacity in the \HI\ line could account for the different CO intensities detected towards regions of similar integrated \HI\ intensity.  Several studies of \HI\ emission and absorption along sight lines passing through the LMC have been conducted \citep{Dickey:94,Mebold:97,Marx:00}, indicating widespread cool \HI\ gas with $T$=30--40 K and peak optical depths of $\tau$=0.5--1.5.  In particular, three of the lines of sight reported by \citet{Dickey:94} show bright CO emission: they are towards 0521-699, 0539-697, and 0540-697, and show peak \HI\ optical depths of 0.91, 1.68, and 1.39 respectively.  Since the \HI\ column density is underestimated by a factor of roughly $\tau/[1-\exp(-\tau)]$ \citep[e.g.,][]{Dickey:03}, towards optically thick sight lines it should be corrected upward by a factor of 1.3--2.  This is in fact an overestimate because opacity in the line wings is much less than at the line center (see e.g., Fig.~2 of \citealt{Marx:00}).  Corrections of this magnitude would not significantly affect the trends seen in Figure~\ref{fig:corrmom}.  

We also note that peak values of $T_b$(\HI) reach values of $T_{\rm max} \sim 100$ K (Fig.~\ref{fig:corrpk}); cool \HI\ at temperatures of $\sim$30 K (as inferred from previous absorption studies) would be a minority constituent of the total \HI\ brightness temperature, and thus not a major factor in the 2-D analysis considered here.

\subsubsection{Variations in Gas Thickness}\label{disc:hz}

The poor correlation between CO and \HI\ intensities could arise from a decoupling of volume and surface densities as a result of variations in the gaseous scale height.  The thickness of the gas layer in a galaxy is related to both the pressure support of the gas and the gravitational force acting on it.  An increase in the gas scale height is most easily achieved by either increasing the velocity dispersion of the gas or decreasing the gravitational force on it.  Consequently, if a thicker gas disk were responsible for lowering the volume density in \HI-rich, CO-poor regions, one would expect these regions to exhibit higher \HI\ velocity dispersion and/or lower stellar surface density.  Fig.~\ref{fig:rmolstar}(a), however, shows no tendency for low H$_2$/\HI\ ratios to be associated with weak stellar gravity (low $\Sigma_*$).  Fig.~\ref{fig:rmolstar}(b) shows a hint of an anti-correlation between $R_{\rm mol}$ and the \HI\ velocity dispersion, although the trend is weak.  If real, such an anti-correlation would be consistent with a lowering of the volume density due to increased turbulent support of the \HI\ layer.  Consistent with this idea, we noted in Figure~\ref{fig:corrgint2}(b) an apparent decrease in CO detectability at larger values of $\sigma_v$(\HI).

On the other hand, turbulence would be expected to simultaneously {\it increase} local volume densities by generating small-scale density structure (as discussed further in \S\ref{disc:warmhi} below), possibly leading to increased CO emission.  An energy source for the turbulence that thickens the gas disk is also needed, which could be either massive star formation or galactic dynamics \citep{Elmegreen:04}.  The abundance of supershells and the lack of strong density wave structures in the LMC suggests that star formation should dominate, but star formation is expected to correlate with higher, not lower, gas density.  Adding to the ambiguity are the possible contributions of cosmic ray, magnetic field, and external pressures.  These factors may not be negligible, especially given the interaction of the LMC with the Galactic halo, and their influence on the gas layer thickness is poorly understood.

\subsubsection{Variations in Density Contrast}\label{disc:warmhi}

Two regions of similar average \HI\ column density could have very different density structures when observed at higher resolution.  A high degree of clumping, leading to larger density contrasts, could increase the amount of material at high enough densities to radiate CO line emission, while still maintaining a fixed column density when averaged over a telescope beam.  Conversely, a more uniform distribution of material could lead to similar average column densities but with relatively little high-density gas.

A straightforward way to model density contrasts is to consider the probability distribution function (PDF) of the volume density, which in the presence of turbulent forcing \citep[e.g.,][]{Wada:01} approaches a log-normal function,
\begin{equation}
f(\rho)\,d\rho = \frac{1}{\sigma\sqrt{2\pi}}\exp\left[-\frac{\ln(\rho/\rho_0)^2}{2\sigma^2}\right]\,d\ln\rho\;.
\end{equation}
\citet{Wada:07} show that the width $\sigma$ of the PDF depends on the ratio of the volume averaged density $\left<\rho\right>$ to the characteristic density $\rho_0$,
\[\frac{\sigma^2}{2} = \ln\left(\frac{\left<\rho\right>}{\rho_0}\right)\;.\]
They therefore derive the mass fraction of gas above a critical density $\rho_c$ to be
\begin{equation}
f_c=\frac{1}{2}\left\{1 - {\rm erf} \left[\frac{\ln(\rho_c/\!\left<\rho\right>)-\sigma^2/2}{\sigma\sqrt{2}}\right]\right\}\;,
\end{equation}
which increases strongly with $\sigma$.  Adopting an average gas density $\left<\rho\right> \sim 0.05$ \Msol\ pc$^{-3}$ or $n_{\rm H} \sim 2$ cm$^{-3}$ for the LMC (using a surface density of $\Sigma_{\rm H} \sim 10$ \Msol\ pc$^{-2}$ and disk thickness of 200 pc, e.g. \citealt{Kim:99}), roughly 17\% of the gas mass will have densities exceeding 100 cm$^{-3}$ for $\sigma$=2, but only 3\% for $\sigma$=1.5.  Therefore, if molecular clouds form primarily as a result of turbulent compression, their existence could be determined primarily by variations in $\sigma$, and high column density would be necessary but not sufficient for high volume density.  For isothermal turbulence, the width $\sigma$ of the PDF is a simple function of the rms Mach number \citep{Padoan:02}, although such a relation may be inapplicable to the ISM as a whole \citep{Wada:07}.  

On the other hand, attributing changes in the fraction of dense gas to the effects of turbulence would lead one to expect larger \HI\ line widths in CO-detected regions, contrary to our results (\S\ref{sec:hidisp}).  A more plausible explanation is a significant component of warm, diffuse gas, which forms a secondary peak distinct from the log-normal portion of the density PDF \citep{Wada:01,Wada:07}.  Despite its low density, the WNM can make a large contribution to the column density averaged over a beam by virtue of its high volume filling factor: in the simulations of \citet{Wada:07}, 80--90\% of the volume is occupied by low-density gas.  In addition, WNM profiles will be broader because of the much higher temperature of the WNM (factor of $\sim$100 higher than CNM, corresponding to a factor of $\sim$10 increase in thermal linewidth).  It is therefore plausible that regions of relatively high \HI\ column density but lacking in CO emission have a higher proportion of WNM gas, which would also account for the observed anti-correlation between CO intensity and \HI\ velocity dispersion.  Of course the physical mechanisms governing the CNM/WNM phase balance would still need to be identified; these would have to be more complex than a simple dependence on \HI\ column density.

Changes in density contrast in the ISM should lead to observable changes in \HI\ line brightness when observing at finer spatial resolution.  If atomic and molecular gas are part of a continuum in gas density, one might also detect analogous signatures by comparing CO with high-density molecular line tracers such as HCO$^+$ and HCN, since the width of the density PDF will affect that fraction of material able to emit strongly in those lines \citep[e.g.,][]{Ballesteros:02}, although abundances and interstellar chemistry will also affect their strength.  Future observations should be able to test these possibilities.

\begin{figure}
\includegraphics[height=9cm]{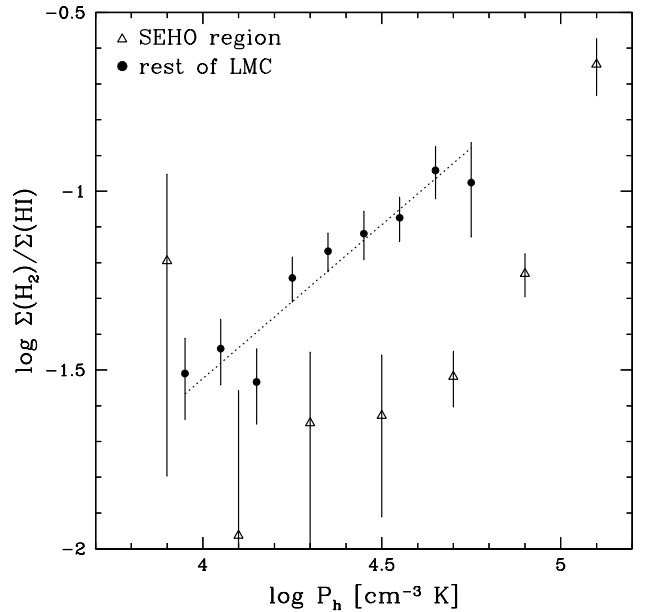}
\caption{
Ratio of molecular to atomic gas surface density as a function of midplane hydrostatic pressure, averaged over all pixels in logarithmically spaced bins of $P_h$.  The open triangles correspond to the southeastern region (SEHO) outlined in Figure~\ref{fig:maps}, while the filled circles (to which a power law of slope 0.86 has been fit, as shown by the dotted line) correspond to the rest of the galaxy.  Note that the highest pressures are only found in the SEHO.  Error bars represent the standard error in the mean for each bin, and non-detections have been included in the bin averages.
\label{fig:binned}}
\end{figure}

\subsection{Molecular to Atomic Ratio and Pressure}\label{disc:press}

Equilibrium analysis of H$_2$ self-shielding suggests that the column density $N_{\rm tr}$ of the \HI\ envelope surrounding a molecular cloud should scale approximately as
\begin{equation}
N_{\rm tr} \propto \left(\frac{n_{\rm gas}R_{\rm H}}{G_{\rm surf}}\right)^{-1.4}
\end{equation}
where $n_{\rm gas}$ is the mean gas density, $R_{\rm H}$ is the rate constant for H$_2$ formation on grains in units of cm$^3$~s$^{-1}$ (depending primarily on metallicity), and $G_{\rm surf}$ is the H$_2$ photodissociation rate at the cloud surface \citep{Federman:79}.  As the ambient pressure (and thus cloud density) increases in the central regions of a galaxy, due to the deeper gravitational potential, molecular cloud envelopes become thinner and the H$_2$/\HI\ ratio ($R_{\rm mol}$) increases \citep{Elmegreen:89}.  This trend has been confirmed in spiral galaxies with strong CO emission \citep{Wong:02,Blitz:06}.  Empirically, the correlation between hydrostatic pressure and the CO/\HI\ ratio arises because the latter is a strong function of radius and also correlates well with the variation in the stellar surface density.  However, in the case of the LMC, many of the massive CO complexes are found well outside the region of high stellar density (the bar).  As indicated by Figure~\ref{fig:rmolstar}(a), no significant correlation between $R_{\rm mol}$ and $\Sigma_*$ is apparent. 

Figure~\ref{fig:rmolstar}(d), on the other hand, does show an apparent separation between detections and upper limits when comparing $R_{\rm mol}$ and $P_h$.  To investigate this further, we plot in Figure~\ref{fig:binned} the average values of $R_{\rm mol}$ in logarithmically spaced bins of $P_h$.  Both detections and non-detections have been included in the averaging, and we have separated the southeastern region of bright \HI\ [dubbed the SE \HI\ overdensity (SEHO) by \citet{Nidever:08}], as defined in Figure~\ref{fig:maps}, from the rest of the galaxy.  Excluding the SEHO, we find a good correlation between $R_{\rm mol}$ and $P_h$, with a best-fit power-law slope of 0.86.  The slope is similar to that found by \citet{Wong:02} and \citet{Blitz:06}, although the agreement may be misleading since both earlier studies only considered CO detections.  Indeed, the correlation can largely be attributed to the increasing CO detection rate with increasing $I_{\rm HI}$ seen in Figure~\ref{fig:corrmom}, since most of the variation in $P_h$ results from variation in $\Sigma_{\rm HI}$ [compare Fig.~\ref{fig:rmolstar}(c) with Fig.~\ref{fig:rmolstar}(d)].  Thus, while there is clearly a relationship between $R_{\rm mol}$ and $P_h$, we prefer to attribute it to the fact that significant \HI\ intensity is a prerequisite for CO emission, rather than any underlying relationship involving the stellar potential or turbulent gas pressure.

The additional implication of Fig.~\ref{fig:binned}, that $R_{\rm mol}$ may behave differently in the SEHO region, is noteworthy given that this region is likely experiencing external ram pressure due to the LMC's motion in the Galactic halo, which may be compressing the gas disk \citep{deboer:98}.  X-ray spectra in this region suggest thermal pressures of $\sim 10^5$ cm$^{-3}$ K \citep{Blondiau:97}.  For comparison, equation~(\ref{eqn:hydro}) gives a fiducial estimate for $P_h$ in the central regions of the LMC (using the surface densities in Figure~\ref{fig:radprof}) of $P/k \sim 10^4$ cm$^{-3}$ K, comparable to estimates for the solar neighborhood.  Clearly, further examination of this and other interacting galaxy environments is needed to better understand how $R_{\rm mol}$ is governed.

Differences in the relationship between $P_h$ and $R_{\rm mol}$ in the LMC as compared to molecule-rich spirals probably indicate that $P_h$ responds more sensitively to atomic gas in the former case and molecular gas in the latter case.
The expression for $P_h$ given in equation~(\ref{eqn:hydro}) can be written as
\begin{equation}
P_h = \frac{1}{2}\Sigma_{\rm gas}\sigma_g\left[\frac{\kappa}{Q_*} + \frac{\kappa}{Q_g}\right] = \frac{\Sigma_{\rm gas}\sigma_g\kappa}{2 Q_{\rm eff}}\,
\end{equation}
where 
\[Q_* \equiv \frac{\kappa\sigma_*}{\pi G\Sigma_*}, \quad
Q_g \equiv \frac{\kappa\sigma_g}{\pi G\Sigma_{\rm gas}}, \quad {\rm and} \quad
Q_{\rm eff} \equiv \left(\frac{1}{Q_*} + \frac{1}{Q_g}\right)^{-1}\;.\]
Here $Q_{\rm eff}$ is an approximate gravitational instability parameter for a combined disk of gas and stars \citep{Wang:94}.  Note that $Q_{\rm eff}/\kappa$ is a measure of the disk dynamical timescale \citep[e.g.,][]{Krumholz:05}, so $P_h = \Sigma_{\rm gas}\sigma_g/(2t_{\rm dyn})$.  (The equivalent expression $P_h = \rho_g\sigma_g^2$ then yields a gas scale height of $h_g \sim \Sigma_{\rm gas}/\rho_g \sim \sigma_g t_{\rm dyn}$, as expected.)  In the inner disks of CO-bright spiral galaxies, $\Sigma_{\rm gas}$ is dominated by molecular gas, $\sigma_g$ is relatively constant, and the dynamical timescale $Q/\kappa$ is expected on theoretical grounds to be proportional to the star formation timescale \citep{Elmegreen:02,Krumholz:05}, which for these regions is also fairly constant \citep{Wong:02,Leroy:08}.  Thus $P_h$ scales linearly with the H$_2$ surface density and with $R_{\rm mol}$ (given the saturation of the \HI\ column density near 10 $M_\odot$ pc$^{-2}$, \citealt{Leroy:08}).  In the case of the LMC, however, most of the gas mass is in \HI, and furthermore the \HI\ column density is generally not saturated, especially when measured locally.  These conditions will ensure that $P_h$ correlates primarily with $\Sigma_{\rm HI}$, which results in a much weaker relationship with $R_{\rm mol}$.

Finally, attempts to predict $R_{\rm mol}$ based on equilibrium values of surface density and/or pressure may be frustrated if, as may be the case in molecule-poor environments, the time to reach equilibrium H$_2$ abundances exceeds the lifetime of clouds against dissociation.  Such a situation has been suggested by \citet{Leroy:07} in relation to the SMC.  Under such conditions no equilibrium would be obtained, and the formation of H$_2$ may be primarily dependent on local conditions such as dust abundance.  Future comparisons of the H$_2$ abundance in the LMC, as measured by UV absorption towards background sources, with other diagnostics of interstellar conditions could be used to test this hypothesis.

\section{Summary and Conclusions}

In this paper we have examined correlations between CO integrated intensity and \HI\ integrated intensity, peak brightness temperature, and linewidth across the LMC.  Our main results can be summarized below:
\begin{enumerate}
\item Large \HI\ integrated intensities and peak brightnesses are prerequisites for detection of CO, with the detection fraction rising steadily for larger values of both quantities, though never reaching 100\% on spatial scales of $\sim$40 pc.
\item On the other hand, neither condition is sufficient to guarantee CO detection, and the quantitative correlations of both quantities with CO intensity are poor.
\item The large range in CO intensities at the highest \HI\ intensities persists even when the data are spatially averaged on $>$200 pc scales, suggesting that local conversion of atomic to molecular gas is not the principal reason for the scatter.
\item The \HI\ linewidth shows no clear correlation with CO intensity, suggesting that molecular clouds do not form directly from the collision of \HI\ clouds.  Decomposing the \HI\ profiles into up to two Gaussian components and identifying the component more closely associated in velocity with CO actually suggests an {\it anti-correlation} between \HI\ linewidth and CO intensity.  This may pose a problem for turbulent models for forming molecular clouds, although it may also reflect a time delay between the initial cloud compression and the appearance of widespread CO emission.
\item While our analysis does not investigate possible differences between CO and \HI\ spectral profiles, we find from our simple Gaussian decomposition that the CO is almost invariably associated with the highest brightness \HI\ component, irrespective of whether this component exhibits ``disk-like'' or ``anomalous'' kinematics.  A full 3-D analysis will be presented in Paper II.
\item We find no clear correlation between the molecular to atomic gas ratio, as traced by $I_{\rm CO}/I_{\rm HI}$, and the hydrostatic pressure $P_h$, when examined on a point-by-point basis.  Since this comparison is problematic because both quantities involve $I_{\rm HI}$, we also search for correlations between $I_{\rm CO}/I_{\rm HI}$ and the stellar surface density and gas velocity dispersion, but again find no significant trends.  That the LMC does not follow the trends observed in CO-bright spiral galaxies may be a consequence of its relatively small molecular fraction.
\end{enumerate}

The fact that CO is always associated with significant \HI\ emission, while not surprising, confirms the importance of atomic gas as the raw material for molecular cloud formation.  That strong \HI\ emission is not always associated with CO suggests several possibilities, such as extensive H$_2$ gas not traced by CO, variations in the timescale for CO formation, or a decoupling between surface and volume density as a result of changes in the gas layer thickness or density contrast.  Although none of these possibilities can be ruled out entirely, we suggest that the presence of a low-density WNM which makes a significant contribution to the observed column density along some lines of sight provides a natural explanation for regions exhibiting similar $N$(\HI) but different amounts of CO emission.  In this context, the broadening of the \HI\ linewidth in WNM-dominated regions might account for the weak anti-correlation between linewidth and CO emission.  

A widespread component of low-density \HI\ would also have implications for predicting the molecular to atomic gas ratio, since such a component does not contribute to shielding H$_2$ against photodissociation (see \citealt{Krumholz:09} for further discussion).  Models which attempt to explain the balance between H$_2$ and \HI\ in terms of shielding would need to take this component into account.  In addition, the time is ripe for undertaking a joint analysis of the existing \HI\ emission and absorption data in order to model the large-scale distributions of the WNM and CNM in the LMC.  The simple Gaussian fitting presented in this paper is inadequate for this task, given the complexity of the \HI\ spectral line profiles, and a combination of spatial and spectral decomposition, as has been recently presented by \citet{Kim:07}, appears to be a more promising approach.

Finally, the utility of CO as a tracer of H$_2$ should continue to be investigated with improved data, especially for the Magellanic Clouds.  The ongoing Mopra survey \citep{Ott:08} will map CO across the LMC at an angular resolution of $\sim$45\arcsec, surpassing that of the \HI\ synthesis data.  Further constraints on the amount of H$_2$ not associated with CO is now being provided by maps of far-infrared or sub-millimeter dust emission, or dust extinction in the near- to mid-infrared \citep[e.g.,][]{Imara:07,Dobashi:08}.  We have also emphasized the complementary role played by UV absorption studies.  In addition, further constraints on the physical properties of the CO-emitting clouds, through observations of high-excitation transitions or optically thin isotopologues, will help to clarify the conditions under which CO is easily detected.

\acknowledgments

The survival analysis used in this study was performed with the ASURV
software provided by the Penn State Center for Astrostatistics. T. W.
acknowledges support from the ATNF and Nagoya University during the early
stages of this project, and support from the University of Illinois during
its completion.  J. O. acknowledges support from NRAO, which is operated by
Associated Universities, Inc., under cooperative agreement with the
National Science Foundation.  J. L. P. was supported by an appointment to the
NASA Postdoctoral Program at the Jet Propulsion Laboratory, administered by
Oak Ridge Associated Universities through a contract with NASA.  S. K. was
supported in part by the Korea Science and Engineering Foundation (KOSEF)
under a cooperative agreement with the Astrophysical Research Center of the
Structure and Evolution of the Cosmos (ARCSEC). We thank the referee for a
detailed report which led to significant improvement and clarification of
the text and figures.

\bibliographystyle{apj}
\bibliography{merged}

\end{document}